\documentclass[preprint]{aastex}
\usepackage{epsf}
\usepackage{rotating}
\usepackage{graphicx}
\usepackage{subfigure}
\begin{document}

\title{Phase-Resolved Infrared Spectroscopy and Photometry of V1500 Cygni, and
a Search for Similar Old Classical Novae}

\author{Thomas E. Harrison\altaffilmark{1,2}}

\affil{Department of Astronomy, New Mexico State University, Box 30001, MSC 
4500, Las Cruces, NM 88003-8001}

\email{tharriso@nmsu.edu}

\author{Randy D. Campbell, James E. Lyke}

\affil{W. M. Keck Observatory, 65-1120 Mamalahoa Hwy., Kamuela, HI 96743}

\email{jlyke@keck.hawaii.edu, randyc@keck.hawaii.edu}

\altaffiltext{1}{Visiting Observer, W. M. Keck Observatory, which is operated 
as a scientific partnership among the California Institute of Technology, the 
University of California, and the National Aeronautics and Space 
Administration. NOAO proposal 2006A-0005.}

\altaffiltext{2}{Visiting Astronomer, Kitt Peak National Observatory, 
National Optical Astronomy Observatory, which is operated by the Association 
of Universities for Research in Astronomy, Inc., under cooperative agreement 
with the National Science Foundation. NOAO proposal 2006A-0046.} 

\begin{abstract}
We present phase-resolved near-infrared photometry and spectroscopy of the 
classical nova V1500 Cyg to explore whether cyclotron emission is present 
in this system. While the spectroscopy do not indicate the presence of
discrete cyclotron harmonic emission, the light curves suggest that a
sizable fraction of its near-infrared fluxes are due to this component. 
The light curves of V1500 Cyg appear to remain dominated by emission from 
the heated face of the secondary star in this system. We have used infrared 
spectroscopy and photometry to search for other potential magnetic systems 
amongst old classical novae. We have found that the infrared light curves of 
V1974 Cyg superficially resemble those of V1500 Cyg, suggesting a
highly irradiated companion. The old novae V446 Her and QV Vul have light 
curves with large amplitude variations like those seen in polars, suggesting 
they might have magnetic primaries. We extract photometry for seventy nine 
old novae from the 2MASS Point Source Catalog and use those data to derive 
the mean, un-reddened infrared colors of quiescent novae. We also extract
$WISE$ data for these objects and find that forty five of them were detected.
Surprisingly, a number of these systems were detected in the $WISE$ 22 $\mu$m 
band. While two of those objects produced significant dust shells (V705 Cas 
and V445 Pup), the others did not. It appears that line emission from their 
ionized ejected shells is the most likely explanation for those detections. 
\end{abstract}
\clearpage

\section{Introduction}

Classical novae (CNe) eruptions are thermonuclear runaways (TNR) on the 
surface of a white dwarf that has been accreting material from a
low mass companion for millennia. Townsley \& Bildsten (2004) show that for 
systems with mass accretion rates of $\dot{M}$ = 10$^{\rm -8}$ to 
10$^{\rm -10}$ M$_{\sun}$ yr$^{\rm -1}$, ignition of the TNR can occur once 
the accumulated envelope on the white dwarf exceeds 
10$^{\rm -5}$ M$_{\sun}$. The resulting eruption can surpass the Eddington 
limit, and eject $\approx$ 10$^{\rm -4}$ M$_{\sun}$ of nuclear processed 
material at high velocity. The eruptions of CNe provide tests of our 
understanding of TNRs, the nucleosynthesis that occurs within the 
burning layers (e.g., Starrfield et al. 2009), and the factors that drive 
the outburst luminosity and shell ejection process (e.g., Shaviv 
2001).

All CNe are cataclysmic variables (CVs). CVs can be separated
into two broad classes: magnetic, and non-magnetic. Magnetic CVs harbor
white dwarfs that have field strengths of B $>$ 1 MG, while the field 
strengths for the white dwarfs in non-magnetic CVs are believed to 
be considerably smaller (though see Warner 2004). In addition, magnetic
CVs are further subdivided into two subclasses: polars, and 
``Intermediate Polars'' (IPs). IPs are believed to have 
lower magnetic field strengths (B $\lessapprox$ 7 MG), than polars. IPs 
are identified by having coherent periodicities that are shorter than their 
orbital periods, believed to originate from processes occurring 
near the magnetic poles of the rapidly rotating white dwarfs in these 
systems. In polars, the stronger
field captures the accretion stream close to the secondary star and
funnels it onto the magnetic poles of the white dwarf. Both the primary
and secondary stars in polars are phase-locked to rotate at the orbital 
period.

That there are magnetic CVs among the old novae is already well known with 
such famous CNe such as DQ Her, V603 Aql, and GK Per being classified as IPs. 
Whether there are true polars among the CNe has not yet been fully answered. 
The best case
for the latter is V1500 Cyg. Stockman et al. (1988) showed that V1500 Cyg
exhibits strong, orbitally modulated circular polarization that they suggested
comes from high-harmonic cyclotron emission. V1500 Cyg is slightly 
asynchronous, so it is not currently a true polar, but Schmidt \& Stockman 
(1991) suggest it will return to synchronism within $\sim$ 150 yrs.  
Another likely highly magnetic CN is V2214 Oph. 
Baptista et al. (1993) compare its visual light curve to the proto-type polar 
AM Her (Crampton \& Cowley 1977, Campbell et al. 2008a). The similarity is 
striking, and it is difficult to envision how such a light curve could be 
produced by a source other than cyclotron emission. 

V1500 Cyg was a remarkable CN, having one of the largest outburst 
amplitudes ($\Delta$m $\sim$ 19.2 mag, Warner 1985), and one 
of the most rapid declines from visual maximum ever observed ($t_{\rm 3}$ = 
3.7 d; Downes \& Duerbeck 2000, Strope et al. 2010). Expansion parallax
measures give a distance of 1.5 kpc (Slavin et al. 1995) to V1500 Cyg, and 
thus an absolute magnitude at maximum of M$_{\rm V_{max}}$ = $-$10.7. The 
only other CN that rivals this luminosity is CP Pup, an object that Warner 
(1985) shows has acted in a similar fashion to V1500 Cyg. Both of these 
objects appear to have had outbursts that were much more luminous than that 
of GK Per (M$_{\rm V_{max}}$ = $-$9.4), the most luminous CN for which a 
high precision parallax has been derived (Harrison et al. 2013a). 

V1500 Cyg is also unusual in that it is only one of two CNe that have been 
observed to have a dramatic pre-outburst rise, brightening by eight magnitudes 
in the eight months leading up to its eruption (Collazzi et al. 2009). Collazzi 
et al. also show that V1500 Cyg is one of five CNe (of 22) that have remained 
much more 
luminous ($\sim$ 10$\times$) after outburst, compared to their pre-eruption 
quiescent levels. This suggests that the accretion rate in V1500 Cyg
remains much higher than observed for typical CNe, and thus, as discussed
by Wheeler (2012), such objects are excellent progenitor candidates
for Type I supernovae.

The question is whether the presence of a strong magnetic field played any
role in the unusual outburst (and current) properties of V1500 Cyg. 
Theoretically, it is unclear if strong magnetic fields play any role in 
shaping the outburst of CNe (Livio et al.  1988, Nikitin et al. 2000), so the 
only way to know for sure is to identify and characterize more highly magnetic 
CNe. We have obtained both infrared spectra 
and photometry of V1500 Cyg, and a small set of other old CNe to search for 
behavior that we can associate with the presence of a magnetic white dwarf 
primary. In the next section we describe our observations, in Section 3 we 
present the infrared spectroscopy and photometry for the program objects, in 
Section 4 we discuss our results, and draw our conclusions in Section 5.

\section{Observations}

In the following we present data from three separate observing runs using 
the Near Infrared Camera (``NIRC'') on Keck 1, the Simultaneous Quad Infrared
Imaging Device (``SQIID'') on the KPNO 2.1 m, and the Infrared Imager 
(``IRIM'') on the KPNO 2.1 m. In addition, we perform a new extraction of 
the $JHK$ photometry for CNe from the 2MASS (Skrutskie et al.  2006) and 
$WISE$ catalogs to allow us to put our observations in context, as well as 
explore the infrared colors of quiescent CNe.

\subsection{NIRC Spectra}

We used NIRC (Matthews \& Soifer 1994) on 
Keck 1 on the nights of 2006 July 7 and 8 UT. NIRC (now retired) used a 256 
square InSb array, with a pixel scale of 0.15 "/pix. Besides imaging in the 
standard near-infrared filters, NIRC had three different grisms providing 
resolutions of R $\sim$ 60 to 120. These grisms allowed for the simultaneous 
observation of two adjoining infrared photometric bandpasses. We observed 
with the GR150 grism and the ``JH'' filter to obtain spectra covering the 
$J$- and $H$-bands, and with the GR120 grism and the ``HK'' filter to cover 
the $H$- and $K$-bands. Note that simultaneous spectra covering $JHK$ were 
not possible, so separate filter/grism moves were necessary with a subsequent 
set of integrations, to obtain such spectra. The individual integration times 
on the old novae in our project were three minutes, except for HR Lyr where 
two minute exposures were used. Separate integrations were obtained at two 
different positions along the slit to allow for background subtraction. 
The spectra were extracted using the standard routines inside 
IRAF. For wavelength calibration, the spectrum of the planetary nebula 
K 3-60 was used. Observations of A0V stars were obtained to provide telluric 
correction. The stronger H I lines intrinsic to the spectra of the A0V stars 
were patched-over using a simple linear fit, before division into our program 
object data.  An observing log is presented in Table 1. The NIRC spectra of 
four of the old novae are shown in Fig. 1. 

\subsection{SQIID Data}

Infrared photometry for fifteen objects was obtained using 
SQIID\footnote[3]{See http//www.noao.edu/kpno/sqiid/sqiidmanual.html}
on the KPNO 2.1 m. SQIID obtains $JHK$ images simultaneously. The observation 
log can be found in Table 1 for the objects for which light curves were 
generated. To obtain sky-subtracted images, all of the program objects were 
observed using an ABBA pattern, where an image is obtained at an initial
position (``A''), the telescope is then nodded ($\sim$ 20") to a second 
position (``B''), before returning to the first position. Data were 
flatfielded using the normal procedure, and differential aperture photometry 
was performed on the targets and four nearby field stars. The 
photometry, listed in Table 2, has been calibrated using the 2MASS data for 
the field stars. The other targets with data appearing in Table 2 (and that 
were too faint to generate useful light curves) were observed at various 
times during the 2006 July 10 to 15 observing run.

\subsection{IRIM}

IRIM\footnote[4]{See http://www.noao.edu/kpno/manuals/irim/irim.html}, now 
retired, used a 256 square HgCdTe NICMOS3 array. We obtained data in the
$JHK'$ filters. Unlike SQIID, images in each filter had to be obtained 
separately.  Data were flatfielded using the normal procedure, and 
differential aperture photometry was performed on the targets and several 
nearby bright stars. The photometry has been calibrated using the 2MASS data 
for these stars, and is listed in Table 2. These targets (T Aur, V705 Cas, 
V723 Cas, DM Gem, BT Mon, and V400 Per) were observed on 1997 
February 12 and 13.

\subsection{2MASS Photometry}

To examine the IR colors of old CNe, we extracted the $JHK$ photometry of 
these objects from the 2MASS Point Source Catalog (PSC) to allow us to make a 
comparison with our new observations. Hoard et al. (2002) have previously 
completed a search of this catalog for CVs, and presented data for three 
dozen CNe. We felt, however, that a re-extraction of the photometry for CNe 
from the 2MASS data base was warranted to determine if additional CNe might
have been missed in the Hoard et al. survey. We searched the PSC for all old CNe 
that had secure identifications in the on-line version of the Downes et al.
(2001) catalog. We then used the 2MASS survey images to verify that 
the source listed by the PSC was consistent with the optical counterpart of 
the CN (there were a significant number of CNe where this was not the case). 
The 2MASS ID for the correct PSC counterpart is listed, along with the $JHK$ 
photometry, in Table 3. In addition to the objects found in a search of the 
PSC, we extracted photometry for nine CNe for which the PSC does not list a 
source, but where there is clearly an object visible on the 2MASS survey 
images consistent with the optical counterpart. While most of these objects 
are faint, not all of them are (e.g., V825 Sco). The final tabulation of the 
2MASS photometry for 79 CNe is listed in Table 3. All of the objects 
listed here erupted prior to the year 2001, except V2467 Cyg (Nova Cygni
2007). This CN was included because its mid-IR spectrum had been published 
(Helton et al. 2012), and those data provided context for the results discussed 
in Section 4.2 

Table 3 also tabulates published values for the visual extinction for the 
detected CNe. These values came from a wide variety of sources, using a wide 
variety of techniques. The determination of extinction for CNe can be quite 
difficult, with values often ranging by a factor of two or more for the same 
nova. References to the sources for all of these extinction values are 
explicitly noted. Values of A$_{\rm V}$ where multiple sources are cited 
are averages of the published values. An italicized value in the visual 
extinction column indicates an estimate made assuming that all CNe have 
similar colors in quiescence. This latter topic will be fully addressed
in Section 4.1.

\subsection{$WISE$ Data}

Harrison et al. (2013b) found that light curve data extracted from the
$WISE$ ``All Sky Single Exposure Table'' were invaluable in showing that the 
infrared excesses of polars were due to strong,
low-harmonic cyclotron emission. To their surprise, several polars
they examined had large amplitude ($\Delta$m $>$ 0.6 mag) variations in 
the $WISE$ bandpasses. Thus, such data might provide additional insight
into the objects under study here, and thus $WISE$ data for the program CNe 
were extracted and examined when possible. 

The $WISE$ mission (Wright et al. 2010) surveyed the entire sky in four 
wavelength bands: 3.4, 4.6, 12, and 22 $\mu$m. The two short bandpasses
(hereafter, ``W1'' and ``W2'') are quite similar to the two short $Spitzer$ 
IRAC bandpasses (Jarrett et al.  2011). The 12 $\mu$m channel (``W3'', 
$\lambda_{\rm eff}$ = 11.56 $\mu$m), is similar to that of the $IRAS$ 12 
$\mu$m bandpass, while the 22 $\mu$m (``W4''; $\lambda_{\rm eff}$ = 22.09 
$\mu$m) bandpass closely resembles the $Spitzer$ MIPS 24 $\mu$m channel 
(Jarrett et al. 2011). Given that mid-IR observations of old CNe might
be feasible from the ground, or with $SOFIA$, we decided to extract the $WISE$ 
data for all of the CNe listed in Table 3. Like for the 2MASS extractions 
above, we insured that each detection listed in the $WISE$ All Sky Survey 
Catalog\footnote[5]{See http://wise2.ipac.caltech.edu/docs/release/allsky/expsup/index.html} was consistent with both
the optical candidate, as well as the 2MASS source for each object, by 
examining the actual $WISE$ images. The photometry for these objects is 
listed in Table 4. A significant fraction of the CNe listed in Table 4 
do not appear in the All Sky Survey Catalog. To extract their fluxes, we 
downloaded the $WISE$ images, and used IRAF aperture photometry
to extract their fluxes using nearby, isolated bright stars to calibrate 
their magnitudes (these objects have been identified in Table 4). Due to the 
large pixel sizes inherent to the $WISE$ images, several objects suffer 
from blending issues with a nearby field star. The listed fluxes for those
sources should be used with caution.

\section{Individual Objects}

There are a number of methods to conclusively show the presence of a magnetic
white dwarf in a CV system. Among these are orbitally modulated polarization
(e.g., Stockman et al. 1988), the presence of Zeeman-split absorption lines 
in the photosphere of the white dwarf primary (e.g., Beuermann et al. 2007), 
or the direct detection of emission from cyclotron harmonics using 
spectroscopy.  Examples of the latter have been presented by Campbell et al. 
(2008a,b,c).

Secondarily, one can mount a circumstantial case (as for V2214 Oph) using light 
curves that show ``polar-like'' variations. Given the spacing of the cyclotron 
harmonics, there can be bandpasses where there is little cyclotron emission, 
while an adjacent bandpass is strongly affected. One example is AM Her, where 
the $J$-band is dominated by ellipsoidal variations from the secondary star,
while the $H$ and $K$-bands are dominated by cyclotron emission (see
Campbell et al. 2008a). This is an excellent method that can be applied when 
the CV is too faint for phase-resolved spectroscopy.

The majority of classical novae are generally too faint to obtain phase-resolved 
near-infrared spectroscopy to ascertain whether orbitally modulated cyclotron 
emission is present. However, the low resolution grism mode of NIRC put many 
old novae within reach. To show what the spectra of two known polars look like 
with this instrument, we present the data for DP Leo and EU UMa in Fig. 2. Both 
of these objects are short period polars, having P$_{\rm orb}$ $\approx$  90 min
(Ramsay et al. 2001, Ramsay et al. 2004). As of yet, there is no estimate of 
the magnetic field strength in EU UMa, but DP Leo has been reported to be a 
two-pole polar with B$_{\rm 1}$ = 76 MG and B$_{\rm 2}$ = 31 MG (Cropper \& 
Wickramasinghe 1993). 

At first glance, the NIR spectrum of DP Leo is unusual, showing peculiar, and 
strongly peaked $J$- and $H$-band continua. Such a spectrum superficially 
resembles those of L-dwarfs, except the water vapor features and the $K$-band 
continuum seen in DP Leo do not match such (red) objects. We feel that the best 
match for these features is an M6V. As Fig. 2 demonstrates, the comparison of 
the spectrum of DP Leo to an M6V shows that it has strong excess emission in the
$J$- and $H$-bands. A 31 MG field would have the $n$ = 2 and $n$ = 3 harmonics 
at 1.7 and 1.15 $\mu$m, respectively. This is where the greatest deviations 
from the M6V spectrum occur. Thus, it is obvious that we have detected
cyclotron emission from this object, and this shows the ability of NIRC to
detect this type of emission in our program objects. 

In contrast to DP Leo, the NIRC spectrum of EU UMa does not have obvious 
cyclotron emission features and, like DP Leo, the secondary star of EU UMa 
appears to have a spectral type near M6V (Fig. 2). There are two broad 
emission features located at 1.09 and 1.30 $\mu$m that are consistent with 
the locations of He I (+ H I Pa$\gamma$) and H I (Pa$\beta$), respectively. 
Given the low resolution of NIRC, these features are somewhat broader than 
expected if they were simply due to emission lines from He I and H I. 
The $J$-band light curve of EU UMa shows large variations ($\pm$ 1.5 mag), 
which is quite typical of polars (Fig. 3). The observed minimum in the 
$J$-band light curve occurred at 03:51 UT on 2006 July 12. Extrapolating the 
ephemeris of EU UMa back to the time of the NIRC spectra (2006 July 8), shows 
that the $J$-band minimum would have occurred at 06:46 UT on that date. 
Given that the JH spectra were obtained between 06:22 and 06:39 UT, indicates 
that these data were obtained close to the predicted time of a $J$-band 
minimum. Thus, the weak emission features seen in the $J$-band could be due 
to an unfavorable viewing angle to the cyclotron source and, if so, are 
consistent with the $n$ = 6 and 7 harmonics for a field with B = 14 MG. With 
this field strength, the $n$ = 3, 4, and 5 harmonics would lie
at 1.52, 1.91, and 2.54 $\mu$m, respectively. Unfortunately, our spectrum of 
EU UMa is too poor to surmise whether these weak features exist. However, 
there is no evidence for similar minima in the $H$- and $K$-band light 
curves, and thus the evidence for near-IR cyclotron emission from EU UMa is 
more ambiguous than that seen in DP Leo. Higher quality spectra and light 
curves will be needed to investigate this source further.

\subsection{V1500 Cygni}

V1500 Cyg erupted in 1975, had a large outburst amplitude, and was a very fast 
nova. A summary of the photometric observations of the outburst can be found in
Ennis et al. (1977), while Ferland et al. (1986) present the spectroscopic
evolution. Using photometric observations, Kaluzny \& Semeniuk (1987) assigned 
an orbital period of P$_{\rm orb}$ = 3.35 hr. Polarimetric observations
by Stockman et al. (1988) found that the circular polarization of V1500 Cyg 
varied over a period that was 1.8\% shorter than the photometric period. 
They attributed this shorter period to the underlying rotation of a magnetic 
white dwarf. Schmidt \& Stockman (1991) showed that this rotational period is 
slowly lengthening, and they suggested that V1500 Cyg was an asynchronous polar 
that would return to synchronism in $\sim$ 150 yr. As they noted, (the 
approximately) sinusoidal polarization curves that they observed for V1500 
Cyg are rarely seen in polars. They suggested that the best explanation for 
this behavior is that there is cyclotron emission from two separate poles on 
opposites sides of the white dwarf. The standard model for the behavior of 
circular polarization in polars suggest that it is maximized when the viewer 
is looking along the accretion column, and minimized when the accretion column 
is at right angles to the line-of-sight (when linear polarization would 
dominate, see Bailey et al. 1982). Thus, at the time of zero circular 
polarization, both accretion columns are approximately located near the limb 
of the white dwarf. Schmidt et al. (1995) provide an updated ephemeris for the 
rotation of the white dwarf, where ``magnetic phase 0'' in this ephemeris  
continues to be defined as the time of the positive zero crossing of the 
circular polarization.  

We present the phase-resolved spectra of V1500 Cyg obtained with NIRC in 
Fig. 4. Due to the necessity of observing in either the ``JH'' or ``HK'' bands, 
simultaneous $JHK$ data are not possible. Thus, on the first night, a nearly 
complete (80\% of an orbit) phase-resolved JH data set was obtained, with 
sparser coverage with the HK grism on the second night. The spectra show 
dramatic evolution over an orbit, with weak emission lines near photometric 
phase zero, and very strong emission lines at phase 0.5. The orbital phases 
in this plot were determined using the (minimum light) ephemeris of Semeniuk 
et al. (1995a). The NIRC spectra suggest that near phase 0, the emission line 
region is almost completely out of view. At the epoch of the $JH$ NIRC spectra, 
the photometric and magnetic phases were essentially identical.

As dramatic as the line evolution is over an orbit, the changing slope of
the continuum is also quite apparent. Schmidt et al. 
(1995) modeled phase resolved UV/optical spectroscopy of this source with 
an irradiated M star with T$_{\rm eff}$ = 3,000 K, but that has a hot face 
with a temperature of 8,800 K. In Fig. 4, we have plotted two blackbodies, 
one with a temperature of 8,800 K (blue line), and a cooler one with 
T$_{\rm eff}$ = 3,000 K. It is clear that neither of these matches the 
observed continuum very well at any orbital phase. In fact, the evolution of 
the spectra of V1500 Cyg over an orbit are superficially similar to the phase 
resolved $JHK$ spectra of the polar VV Pup in its high state (Campbell et al. 
2008b).

The light curves of V1500 Cyg are quite amazing (Fig. 5a), with $\Delta$m 
$\sim$ 1.5 mag variations in the near-IR bands. The Schmidt et al. (1995) model
assumed that a very hot (T$_{\rm eff}$ = 90,000 K) white dwarf irradiates
the cool secondary star heating it to the point where it looks like an F star.
Schmidt et al. also estimate that the orbital inclination angle is 
50$^{\circ}$ $\leq$ $i$ $\leq$ 70$^{\circ}$, and list values for the masses
and radii of the two stellar components. We have used the most recent 
release of the Wilson-Divinney (Wilson \& Divinney 1971) code
WD2010\footnotemark[6]\footnotetext[6]{ftp://ftp.astro.ufl.edu/pub/wilson/lcdc2010/}
to calculate model light curves using those parameters. In addition to our 
$JHK$ data, we include the $V$-band light curve of Semeniuk et al. (1995a) 
in our modeling. For this modeling we have set the minima seen in the light 
curves to phase 0 (corresponding to inferior conjunction of the secondary 
star), using just the photometric orbital period from Semeniuk et al. (1995a). 
We address phasing to both the true photometric and magnetic ephemerides, 
below. As shown in Fig. 5a, the model suggested by Schmidt et al. (green 
line) does not come close to reproducing the observed light curves. This is 
not wholly unexpected, as there is 
some evidence that the white dwarf primary in V1500 Cyg is slowly cooling 
(Somers \& Naylor 1999). Thus, we explored models with cooler white dwarfs, 
and a wider range of inclinations. The best fitting models, holding the 
secondary star temperature to 3,000 K, corresponds to an orbital inclination 
of 30$^{\circ}$, and a white dwarf with a temperature of 58,500 K (red curve 
in Fig. 5a). 

It is clear that the irradiated secondary star models cannot
reproduce the very sharp minima in the $JHK$ light curves seen at phase 0. 
Since there are no reports for an eclipse of the primary by the secondary star 
in V1500 Cyg, this cannot be the explanation for the sharp minima we see 
here. It is also quite obvious that the light curves are not the perfect 
sinusoids that are predicted by the light curve modeling. Starting near phase 
0.7, and lasting until phase 0.9, there is a subtle flattening, or excess, 
above the light curve model. 

In Fig. 5b, we phase the $JHK$ light curves to the photometric maximum 
ephemeris from Semeniuk et al. (1995a). With this phasing, neither photometric 
maxima nor minima fall at their expected places. It is interesting to plot the 
location of the times of the magnetic phase zero. The timing of this 
event is indicated by the dashed lines in Fig. 5b. Just before the calculated
time of magnetic phase zero, there is a decline/dip seen in all three near-IR 
light curves. Given the failure of the light curve models to explain the 
morphology of these light curves, and the presence of a dip near magnetic 
phase zero, suggests that this feature is tied to the visibility of the magnetic 
accretion regions.

If we assume that the two accreting magnetic poles are separated by exactly 
180$^{\circ}$ in longitude, there should be a second self-eclipse one half 
phase later ($\phi_{\rm mag}$ = 0.5) than that which occurs at magnetic phase 0. We indicate the 
location of this time in Fig. 5b with a dotted vertical line. There appears
to be a small discontinuity in the light curves at the time of this event. Thus,
it appears that some of the near-IR luminosity is driven by cyclotron
emission from the magnetic accretion regions. It is impossible to fully
quantify this emission, but if we return to the original Schmidt et al. (1995)
irradiated secondary star model, and simply normalize its light curve to
pass through the data at magnetic phase 0, while attempting to match
the light curve minima, we can make an estimate. When this is done, we find
that during the time when the circularly polarized light curve (see Schmidt 
et al.  1991) is dominated by negatively polarized flux, we see a strong 
excess in all three bandpasses. This excess disappears at $\phi_{\rm mag}$ = 0.
The $J$- and $H$-band light curves show a small excess above the model light 
curve during the maxima of the positively circularized flux, but there is no
such excess in the $K$-band during this time.

Given this scenario, the cyclotron emission supplies about 30\% of
the near-IR luminosity of the system at the time of the maximum in the 
negative circularly polarized flux. Schmidt et al. (1995) note that, besides 
the circular polarization, there is little evidence in the optical for 
significant, modulated emission from the magnetic accretion regions. While 
these two results seem contradictory, it appears to be a rather common 
property of normal polars: both ST LMi (Cropper 1987) and AM Her (Michalsky et 
al.  1977) have similar levels of orbitally modulated circular polarization in
the optical as V1500 Cyg, while having near-IR spectra that display large,
discrete, orbitally modulated cyclotron harmonic emission features (see 
Campbell et al. 2008a). By inference, this suggests that the magnetic 
field strength in V1500 Cyg is similar to that of ST LMi and AM Her,
of order B $\sim$ 13 MG. 

V1500 Cyg appears in the XMM Serendipitous Source Catalog (Watson et al. 2009) 
with a total flux in the 0.2 to 12 keV bandpass of 1.13 $\pm$ 0.16 $\times$ 
10$^{\rm -13}$ erg cm$^{\rm -2}$ s$^{\rm -1}$ (observation date 2002 November 
2). We have downloaded the XMM data and reduced it using procedures outlined
in the ``XMM-Newton ABC Guide''\footnote[7]{http://legacy.gsfc.nasa.gov/xmm/doc/xmm\_abc\_guide.pdf}. We modeled the resulting spectrum using 
XSPEC\footnote[8]{http://heasarc.gsfc.nasa.gov/docs/software/lheasoft/xanadu/xspec/manual/manual.html}. The visual extinction to V1500 Cyg is
A$_{\rm V}$ = 1.54 mag (Ferland 1977), and this corresponds to a hydrogen 
column of N$_{\rm H}$ = 2.8 $\times$ 10$^{\rm 21}$ cm$^{\rm -2}$. We ``froze''
this parameter for all of our models. The X-ray spectra of polars
have been modeled with two components: a soft blackbody, and a harder
bremsstrahlung component. For example, Beuermann et al. (2008) model the
$ROSAT$ data of AM Her with $k$T$_{\rm BB}$ = 27 ev, and a much
harder bremsstrahlung component: $k$T$_{\rm brems}$ = 20 keV. Assuming
this scenario for V1500 Cyg, the best fitting two component model 
($\chi^{\rm 2}$ = 6.8) had $k$T$_{\rm BB}$ = 92 ev, and $k$T$_{\rm brems}$ = 
3.3 keV. However, we find that we get an identical quality of fit 
($\chi^{\rm 2}$ = 6.7) if we use $only$ a thermal bremsstrahlung component 
with $k$T$_{\rm brems}$ = 4.2 keV (see Fig. 6). It is obvious that if there 
is a soft blackbody component in V1500 Cyg, it is not as prominent as is 
often seen in the low states of polars (c.f., Beuermann et al. 2008). 
Obviously, it would be useful to have a much longer exposure X-ray 
observation of V1500 Cyg to better unravel the nature of its spectrum.

The distance to V1500 Cyg obtained from nebular expansion parallax methods
is 1.5 kpc (Slavin et al. 1995). The unabsorbed X-ray flux
in the 0.4 to 10 keV bandpass from our single component model is 1.15 $\times$ 
10$^{\rm -13}$ erg s$^{\rm -1}$ cm$^{\rm -2}$. Thus, the X-ray luminosity in
this bandpass is L$_{\rm brem}$ = 3.1 $\times$ 10$^{\rm 31}$ erg 
s$^{\rm -1}$. To put this in context, the high state hard X-ray luminosity of 
AM Her was observed to be L$_{\rm hard}$ = 1.6 $\times$ 10$^{\rm 32}$ erg 
s$^{\rm -1}$ (Ishida et al. 1997). It is interesting to investigate the timing 
of the XMM observations relative to the magnetic phase of the white dwarf: 0.21
$\leq$ $\phi_{\rm mag}$ $\leq$ 0.60. We plot the XMM EPIC PN light 
curve for V1500 Cyg in Fig. 7. The polarization light curve has a minimum at
$\phi_{\rm mag}$ = 0.5, and maxima at $\phi_{\rm mag}$ = 0.25 and 0.75. 
The X-ray light curve of V1500 Cyg shows a sudden rise at the end of
the observation window which is consistent with the onset of the rapid increase
in the optical polarization at this phase. The maximum in the polarized flux
at $\phi_{\rm mag}$ = 0.75 is about twice that of the maximum at 
$\phi_{\rm mag}$ = 0.25 (see Fig. 2 in Schmidt \& Stockman 1991). As shown in 
Ramsay et al.  (2000), the amount of hard X-ray flux is directly correlated 
with the phasing of the optical polarization. Thus, the XMM observations 
almost certainly did not cover the interval of maximum X-ray flux, and the 
true, mean hard X-ray luminosity is almost assuredly several times higher than 
was observed by XMM. Such a luminosity would be consistent with a polar in 
its high state.

\subsection{V2487 Ophiuchi}

V2487 Oph was the first nova to be detected in X-rays prior to its outburst
(Hernanz \& Sala 2002). It has now been shown that this object is a recurrent
nova (RN), with a previously unrecorded outburst in 1900 (Pagnotta et al.
2009).  This classification is consistent with the rapid decline, t$_{\rm 3}$ 
= 8.2 d (Liller \& Jones 1999). Interestingly, V2487 Oph was detected
in hard X-rays by $INTEGRAL$ some four years after outburst with a flux of 
F(17 $-$ 60 keV) = 1.65 $\times$ 10$^{\rm -11}$ erg s$^{\rm -1}$ cm$^{\rm 2}$ 
(Revnitsev et al. 2008). No other RNe appear in this list of $INTEGRAL$ sources. 
All of the distance 
estimates to V2487 Oph are very large: 8 to 27 kpc (Hernanz \& Sala 2002) and 
27 to 48 kpc (Burlak 2008). The distance estimates from Hernanz \& Sala were 
derived by assuming emission from a soft (T$_{\rm eff}$ = 30 eV) blackbody with 
an effective area of a white dwarf. The larger distances are from the maximum 
magnitude, rate of decay (MMRD, ``t$_{\rm 2}$'' or ``t$_{\rm 3}$'') relations. 
It is not clear that an MMRD derived for CNe is appropriate for RNe (as
shown in Harrison et al. 2013a, the various MMRD relations do not work 
especially well {\it for} CNe).  For example, the RN
U Sco had t$_{\rm 3}$ $\sim$ 5 d in its 1999 outburst which peaked at 
m$_{\rm v}$ $\sim$ 7.5 (Munari et al. 1999). Using the MMRD in Downes \& 
Duerbeck (2000), one would estimate M$_{\rm V}$(max) = $-$10.2 for U Sco, and
incorporating its line-of-sight extinction (A$_{\rm V}$ $\sim$ 1.2 mags), 
derive a distance of 20 kpc. Hachisu et al. (2000) argue for distances of 
$\approx$ 4 to 6 kpc for U Sco. Schaefer (2010) has performed an indepth study
of the Galactic RNe, and argues for a distance of 12 kpc to U Sco based on
the properties of its secondary star. He then argues that V2487 Oph is probably
at roughly the same distance as U Sco.

As Fig. 1 shows, the spectrum of V2487 Oph is blue, and has prominent H I and
He I emission lines. There is no hint of a secondary star in these data.
We can estimate the extinction to V2487 Oph by assuming that part of the 
infrared color excess is due to interstellar extinction. Table 2 lists
the mean infrared photometry for V2487 Oph, and assuming that the IR colors
of this RN are similar to those of old CNe (derived below), we calculate a 
value of $\langle$~A$_{\rm V}$ $\rangle$ = 2.2 mag. Schaefer (2010) estimates 
A$_{\rm V}$ = 1.5 $\pm$ 0.6. If, like U Sco, the $JHK$ fluxes of V2487 Oph
are partially contaminated by its secondary star, the extinction value
we derive will be too high. The IRSA and NED extinction calculators estimate 
the total line-of-sight extinction in this direction to be 2.4 and 3.3 mags, 
respectively. This indicates that the larger distances for V2487 Oph (which
put it on the outskirts of the galactic bulge) are probably incorrect. Given 
its peak outburst magnitude of m$_{\rm v}$ = 9, and its slower decay, 
distances of d $\leq$ 8 kpc seem more appropriate. At such distances, it 
would be the most luminous CV in the $INTEGRAL$ database.

The orbital period of V2487 Oph remains unknown. The infrared light curve
presented in Fig. 8 spans 2.9 hr, and there is only a slight brightening
over this interval. There are no large scale variations that would indicate
an irradiated secondary star, or cyclotron emission regions. These results,
along with its large X-ray luminosity, make V2487 Oph an excellent IP
candidate. While several IPs are old CNe, V2487 Oph could be the first
IP that is an RN. [Though it has been suggested that the multi-periodic
variations in the visual light curve of the RN T Pyx indicate the presence
of a magnetic white dwarf primary (Patterson et al. 1998), the existing X-ray 
data do not appear to support this conclusion (Selvelli et al. 2008).]

\subsection{HR Lyrae, OS Andromedae and V373 Scuti}

We obtained NIRC spectra for three other old novae that were accessible 
during our observing run: HR Lyr, OS And, and V373 Sct. HR Lyr was a 
moderate speed nova (t$_{\rm 3}$ = 97 d; Shears \& Poyner 2007) that erupted 
in 1919, OS And was a fast nova (t$_{\rm 3}$ = 25 d; Kato \& Hachisu 2007) 
that erupted in 1986, and V373 Sct was a moderate speed nova (t$_{\rm 3}$ = 
79 d; Strope et al. 2010) that erupted in 1975. HR Lyr remains surprisingly 
bright in the near-IR ($K_{\rm 2MASS}$ = 14.8), while neither OS And or V373 
Sct are visible on the 2MASS images (though Szkody 1994 reports $J$ = 17.8 for 
OS And in 1988). HR Lyr has the bluest continuum of any of the CNe while OS 
And and V373 Sct have much flatter spectra. The H I and He I emission lines in 
all three of these objects are less prominent than those of V1500 Cyg. In 
addition, none of these sources have reported X-ray detections (including a 
5.2 ksec pointed $ROSAT$ observation of OS And), and thus do not appear to be
good cases for magnetic CVs.

\subsection{The Light Curves of V1974 Cygni, V446 Herculis, QV Vulpeculae, 
V Persei and CP Lacertae}

We obtained $JHK$ light curves for five old novae: V1974 Cyg (Fig. 9),
V446 Her (Fig. 10), QV Vul (Fig. 11), V Per (Fig. 12), and CP Lac (Fig. 13). 
As noted earlier, the light curves of polars can show large variations in one 
or more bandpasses, while not showing variations in an adjoining bandpass. Of 
these five novae, the light curves of both V1974 Cyg and V446 Her show 
significant variations over their orbits. The $K$-band light curve of V1974 
Cyg is reminiscent of that for V1500 Cyg, and both the $J$- and $H$-band light 
curves show what appears to be a minimum near the same time as that seen in 
the $K$-band. This same level of variability is also present in the $WISE$
W1 light curve. It is also important to note that Collazzi et al. (2009) 
demonstrate that, like V1500 Cyg, V1974 Cyg has remained much more luminous 
after outburst than it was prior to its eruption. V1974 Cyg 
was observed to be a supersoft X-ray source that was first detected with 
$ROSAT$ some 265 days after outburst (Balman et al.  1998). There do not 
appear to have been any recent X-ray observations of V1974 Cyg.

The light curve for V446 Her shows a brightening in all three bands, followed 
by a sudden drop. The morphology and amplitude of these variations are very 
similar to those seen in the optical light curves of OU Vir (Mason
et al. 2002). Mason et al. explain those light curves by invoking a high 
inclination system (60$^{\circ}$ $\leq$ $i$ $\leq$ $<$ 70$^{\circ}$) with 
a large, optically thick accretion disk hot spot. In V446 Her, however, the 
observed modulation is larger, and better-defined in the $H$- and 
$K$-bandpasses than it is in $J$-band. One would expect that a hot spot would 
have blue colors, and be more prominent at shorter wavelengths. Perhaps
other sources in the system dilute the contribution of the hot spot in the
$J$-band. The W1 and W2 light curves for V446 Her show $\Delta$m = 0.4 mag 
variations over an orbital period, but those data are too sparse to compare with
the $JHK$ light curves. There is a weak (0.0196 counts s$^{\rm -1}$) $ROSAT$ 
source located within 35" of the position of V446 Her. This position is 
within the typical error bars for weak sources in the ``WGA'' catalog 
(White et al. 1995). 

As shown by Honeycutt et al. (2011), V446 Her exhibits what appear to be
rather normal, and quite frequent, dwarf nova eruptions. The pre-outburst
light curve presented by Collazzi et al. (2009) suggests that these events
were probably occurring in the decades prior to the CNe eruption. Because they
lack accretion disks, such outbursts do not occur in polars. Thus, the white
dwarf in V446 Her is not highly magnetic. GK Per is the only other CNe that 
is known to show dwarf nova outbursts (c.f. Bianchini et al. 1982). V446 Her 
is certainly a source that deserves follow-up X-ray observations to determine 
whether it, like GK Per, is an IP.

The SQIID observations of QV Vul were interrupted by clouds, and thus there is 
a gap of nearly 2 hours in its light curve. The orbital period of QV Vul 
remains unknown, but our data suggests that the period is almost certainly 
in excess of two hours. The $J$- and $K$-band light curves show little 
variation, but the $H$-band light curve reveals large amplitude variations, 
consistent with those seen in polars. There is a pointed (2.6 ksec) $ROSAT$ 
observation for QV Vul (obtained in 1995, eight years after outburst) that 
shows a very weak source (S/N $\leq$ 3) at the correct position. The 
combination of large amplitude variations and X-ray emission are suggestive 
of a magnetic CV. Higher quality data, with better temporal coverage are 
needed.

In contrast, the light curves for V Per and CP Lac show no significant 
variations, though both objects are faint, and the photometric data has 
much larger errors. There is a $ROSAT$ pointed observation for CP Lac that 
shows a very weak source, 0.0032 counts sec$^{\rm -1}$, at the correct 
position for this old nova. No such detection exists for V Per.

\section{Discussion}

While there have been numerous claims for old novae having magnetic white 
dwarfs, the majority of these usually suggest that the system is an IP. The 
two best cases for polars remain V1500 Cyg and V2214 Oph. Our phase-resolved 
spectroscopy did not show any evidence for discrete cyclotron features from
V1500 Cyg. Examination of the $JHK$ light curves, however, indicate the
presence of excess, phase-dependent emission that contributes up to 
30\% of the broadband flux when the circular polarization is at its maximum.
The $JHK$ light curves of V1974 Cyg strongly resemble those of V1500 Cyg. This 
nova erupted more recently (1992), and is the most recent of the CNe for which 
we obtained light curves. Thus, irradiation by a hot white dwarf is certainly
a viable interpretation of those data. That both V1500 Cyg and V1974 Cyg have 
remained much more luminous after their outbursts than seen prior to their 
eruptions, 
strengthens the case for V1974 Cyg harboring a highly magnetic white dwarf.  
Two other objects in our sample, V446 Her and QV Vul, had light curves that 
showed large amplitude variations that are similar to the light curves of 
polars, though V446 Her shows dwarf novae outbursts which rules out such
a classification. Both of these objects do appear to be X-ray sources, and
thus new X-ray observations of all these CNe would be useful to explore 
whether they harbor magnetic white dwarfs.

\subsection{The $JHK$ Colors of Quiescent CNe}

To examine whether the current infrared colors of V1500 Cyg are unusual, 
allowing us a quicker way to identify similar systems amongst old CNe, we 
extracted the $JHK$ photometry of CNe in the 2MASS catalog. With these data 
in hand, we can examine the de-reddened colors of old CNe. We present an 
infrared color-color plot in Fig. 14. The mean de-reddened colors of old CNe
are $\langle$($J$~$-$~$H$)$_{\rm o}$$\rangle$ = 0.22 $\pm$ 0.18 and 
$\langle$($H$~$-$~$K$)$_{\rm o}$$\rangle$ = 0.07 $\pm$ 0.10 (note: all recent 
novae and RNe have been excluded 
from this calculation, as have objects with S/N $<$ 3 in their
photometry, along with GK Per and RR Tel, two objects that have IR colors 
dominated by their secondary stars). In Fig. 14 we also plot the position of 
V1500 Cyg at maximum and minimum light (as stars). V1500 Cyg is redder at 
minimum than maximum, but these colors are not especially unusual when compared 
to those of other old CNe. Without the light curves, there would be nothing 
to indicate that V1500 Cyg had unusual photometric properties.

With a mean infrared color relation defined for old novae, we can attempt to 
use it to estimate the visual extinction for CNe that lack such estimates,
as well as compare how well this technique reproduces published values
of A$_{\rm V}$ for these objects. These estimates are the (first) italicized 
values in the final column of Table 3. Szkody (1994) obtained optical and IR 
photometry for a large sample of old novae, and defined the mean de-reddened 
colors of novae in both ($B$ $-$ $V$) and ($V$ $-$ $J$). Using $V$-band 
photometry of the CNe (from Szkody 1994, or Duerbeck \& Seitter 1987), and 
the 2MASS $J$-band photometry, we have estimated extinctions using the 
relation $\langle$($V$~$-$~$J$)$_{\rm o}$$\rangle$ = 0.27 (Szkody 1994), where possible. 
If there is a second italicized value in the final column of Table 3, it 
results from this estimation technique.

It is always dangerous to assume a global correlation for a diverse
group of objects, and thus it is worthwhile to qualitatively assess the
reddenings derived from the near-IR photometry. The orbital periods of the objects in this table
range from 48 hr for GK Per down to 1.5 hr for CP Pup (though see
Diaz \& Steiner 1991, who suggest CP Pup is an IP, and that this is the white
dwarf spin period). Thus, to fill their 
Roche lobes and transfer matter to their white dwarf primaries, the secondary 
stars in these systems must range in size from subgiants, down to M dwarfs. 
As shown in infrared spectroscopic surveys (e.g., Harrison et al. 2004), most 
CVs with P$_{\rm orb}$ $>$ 3 hr show absorption line features attributable to
their secondary stars. For shorter period systems (P$_{\rm orb}$ $<$ 2 hr),
the secondary stars are much more difficult to detect (e.g., Hamilton et al.
2011). Thus, assuming that hot sources completely dominate the $JHK$ colors of 
old CNe should be assumed with caution for longer period CNe.

A cursory inspection of the estimated extinction values presented in Table 3,
however, shows that assuming mean, quiescent infrared colors for old CNe
is a useful technique for estimating extinction values. For objects with
high quality photometry, the estimated extinction values are quite similar to
the previous published values, especially if several values were averaged to 
produce the A$_{\rm V}$ measurement. In general, all disk-dominated CVs are 
blue, with their spectra being dominated by emission from the white dwarf and 
accretion disk 
hotspot. In the near-IR, we observe these sources on the Rayleigh-Jeans 
tail of those blackbodies. And thus, if there is no significant contamination
from the secondary star, such a technique is fairly robust. It is important
to note that a great variety of techniques were used to derive the ``known'' 
extinction values and these themselves are prone to significant uncertainties 
(c.f., Lance et al. 1988 for a discussion of the reddening estimates
for V1500 Cyg), or were derived 
assuming some sort of photometric relationship not unlike that used here 
(e.g., Weight et al. 1994, Miroshnichenko 1988). 

The main weaknesses of this technique are twofold: 1) infrared colors are
not very sensitive to low values of extinction, and 2) the secondary star 
could contaminate the quiescent colors of longer period CNe. Examples of the 
first weakness are demonstrated by the estimated A$_{\rm V}$ values for RR Pic, 
V533 Her, and HR Del, objects with well-determined, but low values of 
extinction. If there are near-IR excesses for these three objects, they are 
lost in the noise of the relationship. In contrast, two objects that appear to 
have values of their estimated extinctions that are too large when compared to 
the previously published values, QZ Aur (P$_{\rm orb}$ = 8.6 
hr) and BT Mon (P$_{\rm orb}$ = 8.0 hr), are longer period CNe binaries that 
been shown to have secondary stars which are detectable in visual data 
(Campbell \& Shafter 1995, Smith et al. 1998). Clearly, the secondary stars in 
these two systems must contaminate the $JHK$ photometry at a level sufficient 
to produce the unrealistically large values of their estimated extinctions. 
Alternatively, excessively large values of the estimated reddening for some CNe 
might indicate cases of source confusion, especially if the previously 
published extinction value was determined near outburst.

\subsection{$WISE$ and $IRAS$ Data for CNe}

As shown above, we have extracted light curve data from the $WISE$ ``All
Sky Single Exposure Table'' to compare to our $JHK$ light curves for 
a couple of sources. Given the value of such data for easily detecting
the presence of strong, low-harmonic cyclotron emission (see Harrison et al.
2013b), we decided to extract $WISE$ photometry for all of the CNe
with 2MASS detections. Those data are listed in Table 4. From that photometry 
we constructed a color-color plot using the 2MASS $K$-band magnitudes
and the $WISE$ W1 and W2-band photometry shown in Fig. 15.
In this figure we have plotted the locus of main sequence star colors in green.
A large group of old CNe appear to have colors consistent with main sequence 
stars. While it is possible that this might be due to these objects having 
returned to a deep quiescence, having their IR colors dominated by their donor 
stars, it is probably just as likely that a number of these detections are the 
result of source confusion. The bright, isolated CNe (e.g., DQ Her, V603 Aql, 
HR Del) have very blue (W1 $-$ W2) colors, suggesting that candidates with 
(W1 $-$ W2) $\lesssim$ 0.0 are true CNe remnants.

Surprisingly, there were eight detections of CNe in the 22 $\mu$m
bandpass:  V705 Cas (Nova Cas 1993), V1974 Cyg (Nova Cygni 1992),
V2467 Cyg (Nova Cygni 2007), V1494 Aql (Nova Aql 1999), V445 Pup 
(Nova Pup 2000), GK Per (Nova Per 1901), RR Tel (Nova Tel 1944) and HR Del 
(Nova Del 1967). We plot the SEDs of the first six of these in Fig. 16, and 
HR Del in Fig. 17. For V445 Pup and V2467 Cyg, the 2MASS data was obtained 
prior to their outbursts. Of the recent CNe in this set, both V705 Cas (Evans 
et al. 2005, and references therein) and V445 Pup (Lynch et al. 2001) produced 
significant dust shells that can explain the observed SEDs with their 
strong W3 and W4-band excesses. It is interesting, however, that V2467 Cyg,
V1494 Aql, and V1974 Cyg did not appear to produce dust shells (see
Helton et al. 2012, and references therein). Helton et al. used $Spitzer$
observations to show that the mid-IR spectra of all three CNe were dominated
by emission lines. Presumably, this line emission continues to dominate
the mid-IR SEDs of these three CNe, unless there have been recent, unexpected
episodes of dust formation. That V1974 Cyg shows near-IR photometric variations 
similar to those of V1500 Cyg, suggests that the central source in this 
object remains sufficiently luminous to irradiate its secondary star, and
thus could provide the ionizing radiation necessary to produce strong line 
emission from its ejected shell. Such a conclusion 
is bolstered by the fact that there is a $GALEX$ source (objid = 
6371619514870663991, observation date = 2011 September 11) with an NUV 
magnitude of 18.69 $\pm$ 0.08, at the position of V1974 Cyg. 

As noted above, the secondary star of GK Per dominates its SED from the
optical into the mid-IR (see Harrison et al. 2007). The long wavelength 
excess of GK Per above that of its secondary star (see Fig. 16), culminating 
in the detection in the W4 bandpass, can be explained as either due to 
bremsstrahlung emission (see Harrison et al.  2013b), or perhaps due to 
synchrotron emission as seen in its IP cousins AE Aqr and V1223 Sgr (Harrison 
et al. 2007, 2010). Since the time of the $IRAS$ mission, RR Tel has been 
re-categorized as a symbiotic star with an accretion disk (Lee \& Park 1999), 
containing a bright red giant secondary star, and this component dominates its 
infrared SED (Harrison 1992). An $IRAS$ spectrum of RR Tel (Harrison \& 
Gehrz 1988) showed a strong silicate emission feature. We have not plotted its 
SED due to its non-CNe nature.

Of all of the CNe detected in the W4 band it is HR Del that is the 
most surprising. Geisel et al. (1970, ``GKL'') presented sparse photometry of 
this object near outburst, and about three years later (t = 
+1150 d). They estimated a dust temperature of 300 K for the latter date.
Harrison \& Gehrz (1988), Callus et al. (1987) and Dinerstein (1986) reported 
$IRAS$ detections of HR Del. Re-extraction of the $IRAS$ data (see below) 
finds 25 and 60 $\mu$m detections of this source with flux densities
of 480 $\pm$ 43 and 370 $\pm$ 72 mJy, respectively. We have plotted the
data from GKL, $IRAS$, 2MASS, and $WISE$ in Fig. 17. In
addition, we include $UBV$ photometry from Bruch \& Engel (1994). Since
the time of the GKL observations to present day, HR Del has declined by
3.65 mag in the $V$-band. $UBV$ photometry from 1979 (Drechsel \& Rahe 1980)
showed that HR Del had $V$  = 12.23, while more recent photometry (circa 2008)
finds an orbitally averaged value of $\langle V \rangle$ = 12.45, with
maxima reaching $V$ = 12.25 (Friedjung et al. 2010). The AAVSO visual light 
curve of HR Del shows no changes between 1979 and 1983, the year of the $IRAS$ 
mission. Recent AAVSO $V$-band photometry finds no obvious changes since 2008. 
HR Del might have been slightly brighter in 1983 as it is currently,
but not significantly so. In contrast to V1974 Cyg, Collazzi et al. (2009)
found that the pre- and post-eruption brightnesses of HR Del were very
similar.

Comparison of the GKL data set to the $IRAS$ and $WISE$ data suggest a
dust shell that has dimmed and cooled in the intervening time. Strangely, 
however, there does not seem to have been an appreciable change in temperature 
of this putative dust shell between the epochs of the $IRAS$ and $WISE$ 
missions. Both
data sets can be fitted with blackbodies having temperatures T$_{\rm bb}$
$\sim$ 125 K. Perhaps like the dust-free CNe observed by $Spitzer$, it is
line emission that is producing the mid-IR excess in HR Del. Such a
scenario remains viable as, according to Friedjung et al. (2010), the
ultraviolet luminosity of HR Del has changed little over the last thirty 
years. They suggest that continued thermonuclear burning is the best 
explanation for this prolonged period of activity. It would be useful 
to get spectra to ascertain the nature of the mid-IR excess of HR Del.

Harrison \& Gehrz (1988, 1991, 1992, 1994) published a series of papers
on $IRAS$ observations of CNe and concluded that the detection
rate was of order 40\% away from the Galactic Center. In the intervening
decades, the $IRAS$ ``1D'' scan co-addition software SCANPI\footnote[9]{http://scanpiops.ipac.caltech.edu:9000/applications/Scanpi/} used in those surveys
has been markedly improved. Input of the list of objects listed in Table 3 
into SCANPI, and using the selection criteria of Harrison \& Gehrz (1988), 
results in nine detections in the 12 and/or 25 $\mu$m bandpasses (excluding GQ 
Mus and V341 Nor which were in outburst during the $IRAS$ mission). Of these, 
only the 25 $\mu$m detection of HR Del is consistent with the detection
of the CNe remnant. The other eight detections were all the result of source 
confusion due to a nearby object that is identifiable in the $WISE$ images. 
Thus, of the approximately 160 CNe surveyed with $IRAS$, only HR Del and DQ Her 
(see Harrison et al. 2013a) are confirmed as detections of the quiescent 
novae. Old CNe are {\it not} luminous mid/far-infrared sources.

\section{Conclusions}

We have obtained phase resolved near-infrared photometry and spectroscopy of 
V1500 Cyg to search for cyclotron emission in this object to confirm that 
it is a highly magnetic polar. While evidence for significant cyclotron 
emission exists, it is impossible to make secure estimates of its contribution 
to the overall flux as it depends on the model used for the underlying
binary. What is needed is a fuller set of phase-resolved $JHK$ spectra to
separate-out the flux of the component that is modulated at the spin period
of the white dwarf. The data sets for several of the other CNe are intriguing, 
and suggest that follow-up X-ray observations might be useful in confirming 
whether additional magnetic systems exist among old CNe. That both
V1500 Cyg and V1974 Cyg remain much more luminous now compared to
their pre-outburst levels, and that they have similar light curves, suggests
that V1974 Cyg might also harbor a highly magnetic white dwarf. If this
can be proven, perhaps one way to identify magnetic systems 
is specifically targeting those CNe that show a large difference between
their pre-, and post-outburst luminosities.

The $WISE$ W4 band data for several systems indicates that strong mid-IR line 
emission continues to be emitted by some CNe decades after their outbursts. 
Such emission requires a luminous, ionizing source, and suggests that some CNe 
may continue to have stable nuclear burning occurring on the surfaces of their 
white dwarf primaries. The majority of quiescent old CNe do not have 
significant mid/far-infrared emission.

\begin{flushleft}
{\bf References}\\
Austin, S. J., Wagner, R. M., Starrfield, S., Shore, S. N., Sonneborn, G.,
\& Bertram, R. 1996, AJ, 111, 869\\
Bailey, J., Hough, J. H., Axon, D. J., Gatley, I., Lee, T. J., Szkody, P., 
Stokes, G., \& Berriman, G. 1982, MNRAS, 199, 801\\
Balman, S., Krautter, J., \& Ogelman, H. 1998, ApJ, 499, 395\\
Baptista, R., Jablonski, F. J., Cieslinski, D., Steiner, J. E. 1993, ApJ,
406, 67\\
Beuermann, K., El Kholy, E., \& Reinsch, K.  2008, A\&A, 481, 771\\
Beuermann, K., Euchner, F., Reinsch, K., Jordan, S., \& G\"{a}nsicke, B. T.
2007, A\&A, 463, 647\\
Bianchini, A., Hamazaoglu, E., \& Sabbadin, F. 1981, A\&A, 99, 392\\
Bruch, A. \& Engel, A. 1994, A\&A Supp., 104, 79\\
Burlak, M. A. 2008, Astronomy Letters, 34, 249\\
Callus, C. M., Evans, A., Albinson, J. S., Mitchell, R. M., Bode, M. F., 
Jameson, R. F., King, A. R., \& Sherrington, M. 1987, MNRAS, 229, 539\\
Campbell, R. D. \& Shafter, A. W. 1995, ApJ, 440, 336\\
Campbell, R. K., et al. 2008a, ApJ, 683, 409\\
Campbell, R. K., et al. 2008b, ApJ, 678, 1304\\
Campbell, R. K., et al. 2008c, ApJ, 672, 531\\
Collazzi, A. C., Schaefer, B. E., Xiao, L., Pagnotta, A., Kroll, P., \& L\"{o}chel, K. 2009, ApJ, 138, 1846\\
Crampton, D. \& Cowley, A. P. 1977, PASP, 89, 374\\
Cropper, M., \& Wickramasinghe, D. T. 1993, MNRAS, 260, 696\\
Cropper, M. 1987, ApSS, 131, 651\\
Diaz, M. P., \& Steiner, J. E. 1991, PASP, 103, 964\\
Dinerstien, H. L. 1986, AJ, 92, 1381\\
Downes, R. A., \& Duerbeck, H. W., 2000, AJ, 120, 2007\\
Downes, R. E., Webbink, \& Shara, M. M., Ritter, H., Kolb, U., \&
Duerbeck, H. W. 2001, PASP, 113, 764 \\
Duerbeck, H. W. 1981, PASP, 93, 165\\
Drechsel, H. \& Rahe, J. 1980, IBVS 1811\\
Duerbeck, H. W., \& Seitter, W. C. 1987, Ap\&SS, 131, 467\\
Ennis, D., Beckwith, S., Gatley, I., Matthews, K., Becklin, E. E., Elias, J.,
Neugebauer, G., \& Willner, S. P. 1977, ApJ, 214, 478\\
Evans, A., Tyne, V. H., Smith, O., Geballe, T. R., Rawlings, J. M. C.,
\& Eyres, S. P. S. 2005, MNRAS, 360, 1483\\
Ferland, G. J., Lambert, D. L., \& Woodman, J. H. 1986, ApJS, 60, 375\\
Ferland, G. J. 1977, ApJ, 215, 873\\
Friedjung, M., Dennefeld, M., \& Voloshina, I. 2010, A\&A, 521, 84\\
Geisel, S. L., Kleinmann, D. E., \& Low, F. J. 1970 (``GKL''), ApJ, 161, L101\\
Gilmozzi, R., Selvelli, P. L., \& Cassatella, A. 1994, Mem. Soc. Astron. 
Italiana, 65. 199\\
Hachisu, I., Kato, M., Kato, T., Matsumoto, K., \& Nomoto, K. 2000, ApJ, 534,
L189\\
Hamilton, R. T., Harrison, T. E., Tappert, C., \& Howell, S. B. 2001, ApJ,
728, 16\\
Harrison, T. E., Bornak, J., McArthur, B. E., \& Benedict, G. G. 2013a, ApJ,
767, 7\\
Harrison, T. E., Hamilton, E. T., Tappert, C., Hoffman, D. I., \& Campbell,
R. K. 2013b, AJ, 145, 19\\
Harrison, T. E., Bornak, J. E., Rupen, M. P., \& Howell, S. B. 2010, ApJ, 710
325\\
Harrison, T. E., Campbell, R. K., Howell, S. B., Cordova, F. A., \& Schwope,
Harrison, T. E., Osborne, H. L., \& Howell, S. B. 2004, AJ, 127, 3493\\
A. D. 2007, ApJ, 656, 455\\
Harrison, T. E., \& Gehrz, R. D. 1994, AJ, 108, 1899\\
Harrison, T. E. 1992, MNRAS, 259, 17\\
Harrison, T., \& Gehrz, R. D. 1992, AJ, 103, 243\\
Harrison, T. E., \& Gehrz, R. D. 1991, AJ, 101, 587\\
Harrison, T. E., \& Gehrz, R. D. 1988, AJ, 96, 1001\\
Harrison, T. E. 1989, PhD Thesis, University of Minnesota, Minneapolis\\
Helton, L. A., Gehrz, R. D., Woodward, C. E., Wagner, R. M., Vacca, W. D.,
Evans, A., Krautter, J., Schwarz, G. J., Shenoy, D. P., \& Starrfield, S. 
2012, ApJ, 755, 37\\
Henize, K. G., \& Liller, W. 1975, ApJ, 200, 694\\
Hernanz, M. \& Sala, G. 2002, Sci., 298, 393\\
Hoard, D. W., Brinkworth, C. S., \& Wachter, S. 2002, ApJ, 565, 511\\
Honneycutt, R. K., Robertson, J. W., \& Kafka, S. 2011, AJ, 141, 121\\
Hric, L., Petrik, K., Urban, Z., \& Hanzl, D. 1998, A\&AS, 133, 211\\
Iijima, T., \& Nakanishi, H. 2008, A\&A, 482, 865\\
Iijima, T., \& Esenoglu, H. H. 2003, A\&A, 404, 997\\
Ishida, M., Matsuzaki, K., Fujimoto, R., Mukai, K., \& Osborne, J. P. 
1997, MNRAS, 287, 651\\
Jarrett, T. H., Cohen, M., Masci, F., et al. 2011, ApJ, 735, 112\\
Kaluzny, J. \& Semeniuk, I. 1987, Acta. Astron., 37, 349\\
Kato, M., \& Hachisu, I. 2007, ApJ, 657, 1004\\
Kiss, L. L., G\"{o}gh, N., Vinko, J., F\"{u}resz, Csak, B., DeBond, H., Tomson, 
J. R., \& Derekas, A. 2002, A\&A, 384, 982\\
Lance, C. M., McCall, M. L., \& Uomoto, A. K. 1988, ApJS, 66, 151\\
Lee, H. \& Park, M. 1999, ApJ, 515, 199\\
Liller, W. \& Jones, A. F. 1999, IBVS 4774\\
Livio, M., Shankar, A., \& Truran, J. W. 1988, ApJ, 330, 264\\
Lyke, J. E. 2003, AJ, 126, 993\\
Lynch, D. K., Russell, R. W., \& Sitko, M. L. 2001, AJ, 122, 3313\\
Mason, E., Howell, S. B., Szkody, P., Harrison, T. E., Holtzman, J. A.,
\& Hoard, D. W. 2002, A\&A, 396, 633\\
Matthews \& Soifer, B. T. 1994, in ``Infrared Astronomy with Arrays, the 
Next Generation'', ed. I. McLean, ASSL, 190, 239\\
Michalsky, J. J., Stokes, G. M., \& Sotkes, R. A., ApJ, 216, L35\\
Miroshnichenko, A. S. 1988, Soviet Astron., 32, 298\\
Moraes, M., \& Diaz, M. 2009, AJ, 138, 1541\\
Munari, U., Zwitter, T., Tomov, T., Bonifacio, P., Molaro, P., Selvelli, P.,
Tomasella, L., Niedzielski, A., \& Pearce, A. 1999, A\&A, 347, L39\\
Nikitin, S. A., Vshivkov, V. A., \& Snytnikov, V. N., 2000, AstL, 26, 362\\
Orio, M., Covington, J., \& \"{O}gelman, H. 2001, A\&A, 373, 542\\
Pagnotta, A., Schaefer, B. E., Xiao, L., Collazzi, A. C., \& Kroll, P. 2009,
ApJ, 138, 1230\\
Patterson, J., et al. 1998, PASP, 110, 380\\
Ramsay, G., Cropper, M., Mason, K. O., Cordova, F. A., \& Priedhorsky, W.
2004, MNRAS, 347, 95\\
Ramsay, G., Cropper, M., Cordova, F., Mason, K., Mch, R., Pandel, D., \&
Shirey, R. 2001, MNRAS, 326, L27\\
Ramasay, G., Potter, S., Cropper, M., Buckley, D. A. H., \& Harrop-Allin,
M. K. 2000, MNRAS, 316, 225\\
Revnivtsev, M., Sazonov, S., Krivonos, R., Ritter, H., \& Sunyaev, R. 2008,
A\&A 489, 1121\\
Ringwald, F. A., Naylor, T., \& Mukai, K. 1996, MNRAS, 281, 192\\
Schaefer, B. E. 2010, ApJ Supp., 187, 275\\
Schmidt, G. D., \& Stockman, H. S. 1991, ApJ, 371, 749\\
Schmidt, G. D., Liebert, J., \& Stockman, H. S. 1995, ApJ, 441, 414\\
Schwarz, G. J., et al. 2011, ApJS, 197, 31\\
Sekiguchi, K., Feast, M. W., Fairall, A. P., \& Winkler, H. 1989, MNRAS, 241,
311\\
Selvelli, P., Cassatella, A., Gilmozzi, R., \& Gonzalez-Riestra, R. 2008, 
A\&A, 492, 787\\
Semeniuk, I., Olech, A., Nalezyty, M. 1995a, Acta Ast., 45, 747 \\
Semeniuk, I., De Young, J. A., Pych, W., Olech, A., Ruszkowski, M., \& 
Schmidt, R. E.  1995b, Acta Ast., 45, 365\\
Shafter, A. W., \& Abbot, T. M. C. 1989, 339, L75\\
Shaviv, N. J. 2001, MNRAS, 326, 126\\
Shears, J., \& Poyner, G. 2007, JBAAA, 117, 136\\
Skrutskie, M .F., et al. 2006, AJ, 131, 1163\\
Slavin, A. J., O'Brien, T. J., \& Dunlop, J. S. 1995, MNRAS, 276, 353\\
Smith, D. A., Dhillon, V. S., \& Marsh, T. R. 1998, MNRAS, 296, 465\\
Somers, M. W. \& Naylor, T. 1999, A\&A, 352, 563\\
Starrfield, S., Iliadis, C., Hix, W. R., Timmes, F. X., \& Sparks, W.
M., 2009, ApJ, 692, 1532\\
Stockman, H. S., Schmidt, G. D., \& Lamb, D. Q. 1988, ApJ, 332, 282\\
Strope. R. J., Schaefer, B. E., \& Henden, A. A. 2010, AJ, 140, 34\\
Szkody, P. 1994, AJ, 108, 639\\
Thorstensen, J. R., \& Taylor, C. J. 2000, MNRAS, 312, 629\\
Townsley, D. M., \& Bildsten, L. 2004, ApJ, 600, 390\\
Warner, B. 1995, Cataclysmic Variable Stars (Cambridge: Cambridge Univ. Press),
261\\
Warner, B. 2004, PASP, 116, 115\\
Warner, B. 1985, MNRAS, 217, 1p\\
Watson, M. G., Schr\"{o}der, A. C., Fyfe, D., et al. 2009, A\&A, 493, 339\\
Weight, A., Evans, A., Naylor, T., Wood, J. H., \& Bode, M. F. 1994, MNRAS, 266,
761\\
Wheeler, J. C. 2012, ApJ, 758, 123\\
White, N. E., Giommi, P., \& Angelini, L. 1994, BAAS, 185, 4111\\
Wilson, R. E., \& Divinney, E. J. 1971, ApJ, 166, 605\\
Woudt, P. A., \& Warner, B. 2003, MNRAS, 340, 1011\\
Wright, E. L., Eisenhardt, P. R. M., Mainzer, A. K., et al. 2010, AJ, 
140, 1868\\
Yan Tse, J., Hearnshaw, J. B., Rosenzweig, P., Guzman, E., Escalona, O.,
Gilmore, A. C., Kilmarten, P. M., \& Watson, L. C. 2001, MNRAS, 324, 553\\
Yuan, H. B., Liu, X. W., \& Xiang, M. S. 2013, astro-ph1301.1427\\
Zwitter, T., \& Munari, U. 1996, A\&AS, 117, 449\\
\end{flushleft}
\acknowledgements
This publication makes use of data products from the Wide-field 
Infrared Survey Explorer, which is a joint project of the University of 
California, Los Angeles, and the Jet Propulsion Laboratory/California Institute 
of Technology, funded by the National Aeronautics and Space Administration. 
We acknowledge with thanks the variable star observations from the AAVSO 
International Database contributed by observers worldwide and used in this 
research.
\clearpage

\begin{deluxetable}{lcccc}
\centering
\tablecolumns{5}
\tablewidth{0pt}
%\tablehead{Name & $J$ & $H$ & $K$}
\tablecaption{Observation Log}
\startdata
\hline
\hline
Name & UT Date & Start Time & Stop Time & Instrument \\
\hline
DP Leo    & 2006 07 08 & 05:48 & 06:18 & NIRC \\
DP Leo    & 2006 07 11 & 03:30 & 03:46 & SQIID \\
EU UMa    & 2006 07 08 & 06:22 & 06:56 & NIRC  \\
EU UMa    & 2006 07 12 & 03:22 & 04:57 & SQIID \\
V1500 Cyg & 2006 07 07 & 11:27 & 14:06 & NIRC (JH) \\
V1500 Cyg & 2006 07 08 & 12:10 & 13:03 & NIRC (HK) \\
V1500 Cyg & 2006 07 12 & 08:04 & 11:29 & SQIID \\
V2487 Oph & 2006 07 08 & 07:21 & 07:54 & NIRC \\
V2487 Oph & 2006 07 12 & 04:56 & 07:51 & SQIID \\
HR Lyr    & 2006 07 08 & 11:46 & 12:02 & NIRC \\
OS And    & 2006 07 08 & 13:53 & 14:27 & NIRC \\
V373 Sct  & 2006 07 08 & 09:42 & 10:33 & NIRC \\
V1974 Cyg & 2006 07 13 & 05:52 & 07:36 & SQIID \\
V1974 Cyg & 2006 07 13 & 09:49 & 11:45 & SQIID \\
V446 Her  & 2006 07 11 & 04:36 & 08:47 & SQIID \\
V Per     & 2006 07 16 & 08:59 & 11:51 & SQIID \\
QV Vul    & 2006 07 16 & 04:11 & 08:52 & SQIID \\
CP Lac    & 2006 07 11 & 09:02 & 11:55 & SQIID \\
\enddata
\end{deluxetable}

\begin{deluxetable}{cccc}
\centering
\tablecolumns{4}
\tablewidth{0pt}
%\tablehead{Name & $J$ & $H$ & $K$}
\tablecaption{$JHK$ Photometry of Classical Novae}
\startdata
\hline
\hline
Name & $J$ & $H$ & $K$ \\
\hline
T Aur     & 14.06 $\pm$ 0.02 & 13.72 $\pm$ 0.02 & 13.56 $\pm$ 0.04 \\
V705 Cas  & 14.17 $\pm$ 0.04 & 13.79 $\pm$ 0.03 & 13.55 $\pm$ 0.03 \\
V723 Cas  &  9.06 $\pm$ 0.02 &  9.14 $\pm$ 0.02 &  8.46 $\pm$ 0.02 \\
V1330 Cyg & 15.55 $\pm$ 0.06 & 15.25 $\pm$ 0.06 & 15.20 $\pm$ 0.06 \\
V1500 Cyg & 16.84 $\pm$ 0.10 & 16.51 $\pm$ 0.10 & 16.29 $\pm$ 0.10 \\
V1668 Cyg & $\geq$ 18.5  &  $\geq$ 17.4  &  $\geq$ 17.2 \\
V1974 Cyg & 16.04 $\pm$ 0.06 & 15.75 $\pm$ 0.06 & 15.61 $\pm$ 0.06 \\
DM Gem   & 16.49 $\pm$ 0.04 & 16.41 $\pm$ 0.10 & 16.26 $\pm$ 0.12 \\
V446 Her & 16.89 $\pm$ 0.10 & 16.27 $\pm$ 0.10 & 15.76 $\pm$ 0.10 \\
CP Lac & 15.42 $\pm$ 0.05 & 14.94 $\pm$ 0.05 & 14.87 $\pm$ 0.06 \\
DP Leo & 17.81 $\pm$ 0.10 & 17.36 $\pm$ 0.10 & 16.92 $\pm$ 0.14 \\
BT Mon & 14.72 $\pm$ 0.04 & 14.51 $\pm$ 0.06 & 14.42 $\pm$ 0.06 \\
V2487 Oph & 15.48 $\pm$ 0.05 & 14.95 $\pm$ 0.05 & 14.79 $\pm$ 0.05 \\
V Per\tablenotemark{1} & 17.19 $\pm$ 0.22 & 17.10 $\pm$ 0.15 & 16.84 $\pm$ 0.30\\
V400 Per & 18.24 $\pm$ 0.30 & $>$ 18.1 & $>$ 18.1 \\
EU UMa & 17.43 $\pm$ 0.09 & 16.72 $\pm$ 0.07 & 16.51 $\pm$ 0.08 \\
RW UMi & 18.59$\pm$ 0.12 & 18.69 $\pm$ 0.18 & 18.32 $\pm$ 0.24\\
LV Vul & 14.45 $\pm$ 0.12 & 14.01 $\pm$ 0.19 & 14.20 $\pm$ 0.22 \\
NQ Vul & 16.17 $\pm$ 0.19 & 15.41 $\pm$ 0.20 & 15.11 $\pm$ 0.28\\
QU Vul & $>$ 18.5 & $>$ 17.5 & $>$ 17.3 \\
QV Vul\tablenotemark{1} & 16.24 $\pm$0.15 & 16.13  $\pm$ 0.16 & 16.14 $\pm$0.21 \\
\enddata
\tablenotetext{1}{The errors in the photometry for this object were calculated using the 
standard deviation of the data for the entire light curve (all other error bars are 
those typical for a single measurement).}
\end{deluxetable}
\clearpage
\begin{deluxetable}{cccccc}
\centering
\tablecolumns{6}
\tablewidth{0pt}
%\tablehead{Name & $J$ & $H$ & $K$}
\tablecaption{2MASS $JHK$ Photometry of Classical Novae}
\startdata
\hline
\hline
Name & 2MASS ID & $J$ & $H$ & $K$ & A$_{\rm V}$\tablenotemark{1}\\
\hline
V723 Cas  &01050535+5400402&11.88 $\pm$ 0.02&11.69 $\pm$ 0.02&10.93 $\pm$ 0.02&1.5$^{\rm a}$,$\cdots$\\
GK Per    &03311201+4354154&10.86 $\pm$ 0.02&10.27 $\pm$ 0.02&10.06 $\pm$ 0.02&0.9$^{\rm b}$,$\cdots$\\
QZ Aur    &05283408+3318218&15.83 $\pm$ 0.06&15.29 $\pm$ 0.09&14.97 $\pm$ 0.11&1.7$^{\rm c}$,{\it 3.5}\\
T Aur     &05315911+3026449&14.02 $\pm$ 0.02&13.74 $\pm$ 0.03&13.58 $\pm$ 0.03&1.6$^{\rm c,d,e}$,{\it 1.0}\\
RR Pic    &06353606-6238242&12.46 $\pm$ 0.02&12.40 $\pm$ 0.02&12.25 $\pm$ 0.02&0.1$^{\rm b}$,{\it 0.0}\\
BT Mon    &06434723-0201139&14.40 $\pm$ 0.04&13.96 $\pm$ 0.05&13.72 $\pm$ 0.05&0.6$^{\rm d}$,{\it 2.4} \\
DN Gem    &06545435+3208280&15.43 $\pm$ 0.05&15.24 $\pm$ 0.09&15.26 $\pm$ 0.17&0.3$^{\rm c,e,i}$,{\it 0.0}\\
GI Mon    &07264710-0640297&15.33 $\pm$ 0.05&15.02 $\pm$ 0.08&14.81 $\pm$ 0.12&0.3$^{\rm k}$,{\it 1.5}\\
V445 Pup  &07375688-2556589&12.27 $\pm$ 0.02&11.94 $\pm$ 0.02&11.52 $\pm$ 0.02&1.6$^{\rm l}$,{\it 3.3}\\
HS Pup    &07532487-3138503&16.35 $\pm$ 0.10&15.85 $\pm$ 0.16&15.36 $\pm$ 0.17&{\it 4.6/2.0}\\
HZ Pup    &08032287-2828287&15.91 $\pm$ 0.08&15.64 $\pm$ 0.12&15.66 $\pm$ 0.27&{\it 0.0}\\
CP Pup    &08114606-3521049&14.34 $\pm$ 0.03&14.24 $\pm$ 0.04&14.03 $\pm$ 0.07&0.8$^{\rm d,i}$,{\it 0.6}\\
V365 Car  &11031677-5827249&15.64 $\pm$ 0.07&15.31 $\pm$ 0.10&15.24 $\pm$ 0.19&3.2$^{\rm m}$,{\it 0.5}\\
LZ Mus    &11560927-6534201&14.46 $\pm$ 0.03&13.78 $\pm$ 0.04&13.51 $\pm$ 0.05&{\it 3.7}\\
CP Cru    &12103135-6014471&15.95 $\pm$ 0.08&15.54 $\pm$ 0.12&15.14 $\pm$ 0.17&5.9$^{\rm n}$,{\it 3.5}\\
AP Cru    &12312045-6426251&15.46 $\pm$ 0.05&14.76 $\pm$ 0.07&14.49 $\pm$ 0.10&{\it 3.8}\\
V888 Cen  &13023180-6011362&14.48 $\pm$ 0.04&14.12 $\pm$ 0.02&14.13 $\pm$ 0.06&1.3$^{\rm f}$,{\it 0.2}\\
V842 Cen  &14355255-5737352&14.83 $\pm$ 0.06&14.53 $\pm$ 0.08&14.46 $\pm$ 0.12&1.7$^{\rm g}$,{\it 0.4}\\
CT Ser    &15453907+1422317&16.06 $\pm$ 0.14&15.67 $\pm$ 0.14&15.59 $\pm$ 0.19&0.7$^{\rm c}$,{\it 0.8}\\
X Ser     &16191767-0229295&15.50 $\pm$ 0.06&15.00 $\pm$ 0.07&14.78 $\pm$ 0.09&0.8$^{\rm h}$,{\it 2.5}\\
OY Ara    &16405045-5225479&15.57 $\pm$ 0.08&14.82 $\pm$ 0.10&14.62 $\pm$ 0.09&{\it 1.7},{\it 3.5}\\
V841 Oph  &16593037-1253271&12.32 $\pm$ 0.03&11.95 $\pm$ 0.02&11.81 $\pm$ 0.02&1.5$^{\rm c,i}$,{\it 1.3}\\
V2487 Oph &17315980-1913561&15.36 $\pm$ 0.08&14.88 $\pm$ 0.10&14.41 $\pm$ 0.09&1.4$^{\rm a,j}$,{\it 2.2}\tablenotemark{2}\\
V972 Oph  &17344444-2810353&13.62 $\pm$ 0.06&12.94 $\pm$ 0.08&12.61 $\pm$ 0.07&1.5$^{\rm o}$,{\it 4.2}\\
V794 Oph\tablenotemark{3}&17384924-2250489&14.15 $\pm$ 0.06&13.37 $\pm$ 0.05&13.24 $\pm$ 0.06&{\it 3.1/4.2}\\
V721 Sco  &17422909-3440414&13.69 $\pm$ 0.04&13.31 $\pm$ 0.06&13.20 $\pm$ 0.06&{\it 1.1} \\
V3888 Sgr &17484147-1845347&15.19 $\pm$ 0.11&14.52 $\pm$ 0.11&14.32 $\pm$ 0.12&3.3$^{\rm c,p}$,{\it 3.1}\\
V3964 Sgr &17494261-1723361&15.78 $\pm$ 0.10&14.84 $\pm$ 0.14&14.57 $\pm$ 0.18&{\it 4.9} \\
V825 Sco  &   $\cdots$     &15.32 $\pm$ 0.12&14.38 $\pm$ 0.11&14.11 $\pm$ 0.15&{\it 4.9} \\
V1172 Sgr &17502366-2040298&12.57 $\pm$ 0.04&11.41 $\pm$ 0.04&11.03 $\pm$ 0.03&1.2$^{\rm c}$,{\it 6.8}\\
V977 Sco  &17515035-3231575&14.10 $\pm$ 0.06&13.23 $\pm$ 0.05&13.07 $\pm$ 0.04&{\it 3.8} \\
V4643 Sgr &17544041-2614154&13.18 $\pm$ 0.07&12.59 $\pm$ 0.05&12.27 $\pm$ 0.05&5.2$^{\rm a}$,{\it 3.7}\\
V1178 Sco &17570664-3223053&13.74 $\pm$ 0.07& 13.62$\pm$ 0.11&13.33 $\pm$ 0.08&{\it 2.2}\\
V990 Sgr  &17571823-2819079&12.98 $\pm$ 0.05&11.93 $\pm$ 0.05&11.54 $\pm$ 0.04&{\it 6.4} \\
DQ Her    &18073024+4551325&13.60 $\pm$ 0.03&13.28 $\pm$ 0.04&13.09 $\pm$ 0.04&0.3$^{\rm b}$,{\it 1.4}\\
V533 Her  &18142048+4151221&14.71 $\pm$ 0.03&14.65 $\pm$ 0.05&14.64 $\pm$ 0.06&0.3$^{\rm c,d,e,i}$,{\it 0.0} \\
GR Sgr    &18225850-2534473&14.87 $\pm$ 0.07&14.53 $\pm$ 0.05&14.51 $\pm$ 0.08&{\it 0.2/1.0} \\
V1151 Sgr &   $\cdots$     &15.81 $\pm$ 0.12&14.89 $\pm$ 0.10&14.75 $\pm$ 0.13&{\it 3.8} \\
BS Sgr    &18264672-2708199&12.84 $\pm$ 0.03&12.22 $\pm$ 0.03&12.12 $\pm$ 0.03&{\it 2.1/1.7} \\
FH Ser    &18304704+0236520&15.25 $\pm$ 0.06&15.00 $\pm$ 0.11&14.66 $\pm$ 0.11&2.3$^{\rm c,d,e,i}$,{\it 2.3} \\
V1017 Sgr &18320447-2923125&11.45 $\pm$ 0.03&10.85 $\pm$ 0.03&10.67 $\pm$ 0.03&{\it 2.6} \\
V3645 Sgr &18354936-1841385&13.43 $\pm$ 0.03&12.55 $\pm$ 0.03&12.36 $\pm$ 0.03&{\it 4.0} \\
V827 Her  &   $\cdots$     &16.70 $\pm$ 0.18&16.06 $\pm$ 0.20&$\geq$ 15.9     &{\it 3.9} \\
V522 Sgr  &18480046-2522219&15.02 $\pm$ 0.05&14.60 $\pm$ 0.06&14.60 $\pm$ 0.08&1.2$^{\rm c}$,{\it 0.4} \\
V603 Aql  &18485464+0035030&11.70 $\pm$ 0.02&11.51 $\pm$ 0.03&11.35 $\pm$ 0.03&0.2$^{\rm b,c}$,{\it 0.6}\\
HR Lyr    &18532505+2913377&15.06 $\pm$ 0.04&14.81 $\pm$ 0.07&14.82 $\pm$ 0.08&0.5$^{\rm c,i}$,{\it 0.0}\\
EL Aql    &18560202-0319205&13.41 $\pm$ 0.03&12.55 $\pm$ 0.03&12.23 $\pm$ 0.02&{\it 4.9/6.7} \\
V446 Her$^{\rm bl?}$&$\cdots$&15.17 $\pm$ 0.07&14.61 $\pm$ 0.06&14.50 $\pm$ 0.11&1.4$^{\rm c,d,e}$,{\it 1.9} \\
FS Sct    &18581691-0524051&14.65 $\pm$ 0.04&13.93 $\pm$ 0.04&13.71 $\pm$ 0.04&{\it 3.5/3.6} \\
V1059 Sgr &19015056-1309420&15.61 $\pm$ 0.07&15.33 $\pm$ 0.11&14.84 $\pm$ 0.16&{\it 3.6} \\
QV Vul    &   $\cdots$     &17.35 $\pm$ 0.29&16.27 $\pm$ 0.23&$\geq$ 16.1     &1.0$^{\rm q}$,{\it 0.0}\tablenotemark{2} \\
V1378 Aql &19163547+0343263&16.06 $\pm$ 0.11&15.51 $\pm$ 0.10&15.08 $\pm$ 0.16&{\it 4.4/4.7} \\
V356 Aql  &   $\cdots$     &17.25 $\pm$ 0.26&17.00 $\pm$ 0.39&16.40 $\pm$ 0.38&2.0$^{\rm d}$,{\it 4.4} \\
V1301 Aql &   $\cdots$     &18.80 $\pm$ 0.34&16.92 $\pm$ 0.23&$\geq$ 16.8     &2.8$^{\rm p}$,$\cdots$ \\
V528 Aql  &   $\cdots$     &16.55 $\pm$ 0.19&17.15 $\pm$ 0.35&16.42 $\pm$ 0.33&2.6$^{\rm d}$,$\cdots$ \\
V1494 Aql$^{\rm bl?}$&19230529+0457190&14.85 $\pm$ 0.05&14.37 $\pm$ 0.07&14.13 $\pm$ 0.07&1.8$^{\rm r}$,{\it 2.6} \\
V1370 Aql &19232125+0229262&16.20 $\pm$ 0.09&15.42 $\pm$ 0.11&15.30 $\pm$ 0.18&1.8$^{\rm e}$,{\it 3.0} \\
V1229 Aql &19244452+0414486&16.23 $\pm$ 0.15&15.68 $\pm$ 0.12&15.54 $\pm$ 0.20&1.5$^{\rm d,e,p}$,{\it 2.1} \\
PW Vul    &   $\cdots$     &16.11 $\pm$ 0.15&15.81 $\pm$ 0.16&15.66 $\pm$ 0.20&1.6$^{\rm c,p,q}$,{\it 1.1} \\
V368 Aql  &19263446+0736137&14.53 $\pm$ 0.04&14.15 $\pm$ 0.05&14.03 $\pm$ 0.07&0.8$^{\rm c}$,{\it 1.1}\\
NQ Vul$^{\rm bl}$&19291475+2027596&15.03 $\pm$ 0.05&14.63 $\pm$ 0.05&14.26 $\pm$ 0.08&1.7$^{\rm c,e,s}$,{\it 3.2} \\
DO Aql    &19312591-0625382&16.15 $\pm$ 0.12&15.59 $\pm$ 0.15&15.31 $\pm$ 0.19&{\it 3.3} \\
LV Vul    &19480043+2710173&14.65 $\pm$ 0.08&14.03 $\pm$ 0.07&13.80 $\pm$ 0.08&1.7$^{\rm d,p}$,{\it 3.1} \\
V476 Cyg  &19582445+5337075&16.01 $\pm$ 0.10&15.59 $\pm$ 0.15&15.56 $\pm$ 0.19&0.7$^{\rm d,e}$,{\it 0.6} \\
RR Tel    &20041854-5543331& 7.30 $\pm$ 0.04& 6.08 $\pm$ 0.04& 4.90 $\pm$ 0.01&0.3$^{\rm c}$,$\cdots$ \\
V2467 Cyg &20281249+4148365&15.73 $\pm$ 0.06&15.13 $\pm$ 0.10&14.74 $\pm$ 0.12&4.6$^{\rm a}$,{\it 4.3} \\
V1974 Cyg &20303161+5237513&15.67 $\pm$ 0.07&15.28 $\pm$ 0.10&15.21 $\pm$ 0.16&1.1$^{\rm t,u}$,{\it 0.8} \\
HR Del    &20422035+1909394&12.32 $\pm$ 0.02&12.28 $\pm$ 0.02&12.22 $\pm$ 0.02&0.5$^{\rm c,d,i}$,{\it 0.0}\\
V1330 Cyg &20524536+3559213&15.72 $\pm$ 0.10&15.11 $\pm$ 0.09&15.00 $\pm$ 0.12&{\it 2.1/2.1}\\
V450 Cyg  &20584763+3556304&15.02 $\pm$ 0.06&14.51 $\pm$ 0.07&14.41 $\pm$ 0.08&1.4$^{\rm d}$,{\it 1.5}\\
V2275 Cyg &21030196+4845534&16.25 $\pm$ 0.12&16.03 $\pm$ 0.18&16.51 $\pm$ 0.39&3.1$^{\rm v}$,$\cdots$\\
V1500 Cyg &21113632+4809058&16.13 $\pm$ 0.12&15.49 $\pm$ 0.13&15.38 $\pm$ 0.20&1.5$^{\rm w}$,{\it 2.3}\\
Q Cyg     &21414393+4250290&13.54 $\pm$ 0.03&13.25 $\pm$ 0.03&13.11 $\pm$ 0.03&1.4$^{\rm c,i}$,{\it 0.9}\\
IV Cep    &22043692+5330236&14.96 $\pm$ 0.07&14.63 $\pm$ 0.08&14.30 $\pm$ 0.08&1.9$^{\rm c,d,e,p}$,{\it 2.6}\\
CP Lac    &22154108+5537014&15.03 $\pm$ 0.06&14.73 $\pm$ 0.07&14.81 $\pm$ 0.12&1.5$^{\rm c,d}$,{\it 0.8}\\
DI Lac    &22354848+5242596&13.73 $\pm$ 0.04&13.54 $\pm$ 0.05&13.40 $\pm$ 0.04&1.3$^{\rm c,i}$,{\it 0.5}\\
DK Lac    &22494690+5317192&16.43 $\pm$ 0.15&16.00 $\pm$ 0.21&16.11 $\pm$ 0.25&1.3$^{\rm c,d,e}$,{\it 2.0}\\
V705 Cas  &23414719+5730594&14.88 $\pm$ 0.04&14.54 $\pm$ 0.07&14.28 $\pm$ 0.08&1.7$^{\rm x}$,{\it 2.1}\\
BC Cas    &23511745+6018100&15.44 $\pm$ 0.05&14.96 $\pm$ 0.07&14.58 $\pm$ 0.08&3.7$^{\rm y}$,{\it 3.7}\\ 
\enddata
\tablenotetext{bl}{\scriptsize This object suffers from blending issues.}
\tablenotetext{1}{\scriptsize Italicized values for the visual extinction indicates an estimated 
extinction using the 2MASS observed colors. If a second value is present, this extinction
has been estimated using the mean ($V$ $-$ $J$) color relationship for old novae
derived in Szkody (1994).  }
\tablenotetext{2}{\scriptsize Estimated extinction calculated using the data from Table 2.}
\tablenotetext{3}{\scriptsize Woudt \& Warner (2003) note that this object shows no variability.
The optical/IR colors of this source indicate a reddened K-giant. The 2MASS images 
show a faint source blended on the NW side of the object identified as V794 Oph, but that 
faint source has similar colors to the candidate.}
\begin{flushleft}
{\scriptsize
Published A$_{\rm V}$ sources: $^{\rm a}$Schwarz et al. (2011), 
$^{\rm b}$Harrison et al. (2013a), $^{\rm c}$Weight et al. (1994),
$^{\rm d}$Duerbeck (1981), $^{\rm e}$Harrison (1989), $^{\rm f}$Yan 
Tse et al. (2001), $^{\rm g}$Sekiguchi et al. (1989), $^{\rm h}$Thorstensen 
\& Taylor (2000), $^{\rm i}$Bruch \& Engel (1994), $^{\rm j}$Schaefer (2010),
$^{\rm k}$Gilmozzi et al. (1994), $^{\rm l}$Iijima \& Nakanishi (2008),
$^{\rm m}$Henize \& Liller (1975), $^{\rm n}$Lyke et al. (2003),
$^{\rm o}$Zwitter \& Munari (1996), $^{\rm p}$Miroshnichenko (1988),
$^{\rm q}$Orio et al. (2001), $^{\rm r}$Iijima \& Esenoglu (2003),
$^{\rm s}$Warner (1995), $^{\rm t}$Kato \& Hachisu (2007), 
$^{\rm u}$Austin et al. (1996), $^{\rm v}$Kiss et al. (2002),
$^{\rm w}$Ferland et al. (1977), $^{\rm x}$Hric et al. (1998),
$^{\rm y}$Ringwald et al. (1996)\\}
\end{flushleft}
\end{deluxetable}
\clearpage

\begin{deluxetable}{ccccc}
\centering
\tablecolumns{5}
\tablewidth{0pt}
%\tablehead{Name & $J$ & $H$ & $K$}
\tablecaption{$WISE$ Photometry of Classical Novae}
\startdata
\hline
\hline
Name & 3.4 $\mu$m & 4.6 $\mu$m & 12 $\mu$m & 22 $\mu$m\\
\hline
V723 Cas    &13.39 $\pm$ 0.03&13.33 $\pm$ 0.03&12.23 $\pm$ 0.26& $\cdots$ \\
GK Per      & 9.87 $\pm$ 0.02& 9.76 $\pm$ 0.02& 9.30 $\pm$ 0.03& 7.89 $\pm$ 0.15 \\
QZ Aur      &14.91 $\pm$ 0.05&14.87 $\pm$ 0.10&12.52 $\pm$ 0.28& $\cdots$ \\
T Aur       &13.53 $\pm$ 0.03&13.39 $\pm$ 0.04&11.93 $\pm$ 0.28& $\cdots$ \\
RR Pic      &11.99 $\pm$ 0.02&11.82 $\pm$ 0.02&11.33 $\pm$ 0.07& $\cdots$ \\
GI Mon      &15.08 $\pm$ 0.06&15.03 $\pm$ 0.11& $\cdots$ & $\cdots$ \\
V445 Pup    &10.39 $\pm$ 0.02& 7.30 $\pm$ 0.02& 0.55 $\pm$ 0.01&$-$1.32 $\pm$ 0.01 \\
HS Pup      &15.44 $\pm$ 0.05&15.98 $\pm$ 0.24& $\cdots$ & $\cdots$ \\
HZ Pup      &15.28 $\pm$ 0.05&15.03 $\pm$ 0.10&11.89 $\pm$ 0.21& $\cdots$ \\
CP Pup      &13.78 $\pm$ 0.03&13.65 $\pm$ 0.04&12.39 $\pm$ 0.35& $\cdots$ \\
LZ Mus      &14.46 $\pm$ 0.03&13.75 $\pm$ 0.04&13.51 $\pm$ 0.05& $\cdots$ \\
V888 Cen$^{\rm ex}$ &13.96 $\pm$ 0.20&14.15 $\pm$ 0.34& $\cdots$ & $\cdots$ \\
CT Ser      &15.56 $\pm$ 0.05&15.77 $\pm$ 0.17& $\cdots$ & $\cdots$ \\
X Ser       &14.70 $\pm$ 0.04&14.61 $\pm$ 0.07& $\cdots$ & $\cdots$ \\
V841 Oph    &11.74 $\pm$ 0.02&11.70 $\pm$ 0.03&11.56 $\pm$ 0.34 & $\cdots$ \\
V2487 Oph$^{\rm bl}$&14.13 $\pm$ 0.07 &14.44 $\pm$ 0.12 &$\cdots$ & $\cdots$ \\
DQ Her      &12.91 $\pm$ 0.02&12.76 $\pm$ 0.03&12.26 $\pm$ 0.22 & $\cdots$ \\
V533 Her    &14.48 $\pm$ 0.03&14.43 $\pm$ 0.05& $\cdots$ & $\cdots$ \\
BS Sgr$^{\rm ex}$  &11.19 $\pm$ 0.02&11.25 $\pm$ 0.03&11.17 $\pm$ 0.07& $\cdots$ \\
V827 Her$^{\rm ex}$ &15.47 $\pm$ 0.22&$\cdots$ & $\cdots$ & $\cdots$ \\
V603 Aql    &10.80 $\pm$ 0.03&10.72 $\pm$ 0.03&10.01 $\pm$ 0.29& $\cdots$ \\
HR Lyr      &14.76 $\pm$ 0.05&14.73 $\pm$ 0.07& $\cdots$ & $\cdots$ \\
V446 Her$^{\rm bl,ex}$ &13.21 $\pm$ 0.07&13.37 $\pm$ 0.10&12.16 $\pm$ 0.21& $\cdots$ \\
FS Sct$^{\rm bl,ex}$  &12.86 $\pm$ 0.07&12.82 $\pm$ 0.10& $\cdots$ & $\cdots$ \\
V1059 Sgr$^{\rm bl?,ex}$ &14.81 $\pm$ 0.11&14.77 $\pm$ 0.17&13.12 $\pm$ 0.35& $\cdots$ \\
V1494 Aql$^{\rm ex}$   &13.76 $\pm$ 0.07&13.65 $\pm$ 0.10&12.31 $\pm$ 0.22& 7.234 $\pm$ 0.20\\
V1370 Aql$^{\rm ex}$  &14.79 $\pm$ 0.12&14.73 $\pm$ 0.18& $\cdots$ & $\cdots$ \\
V1229 Aql$^{\rm ex}$   &15.34 $\pm$ 0.07&16.02 $\pm$ 0.37& $\cdots$ & $\cdots$ \\
PW Vul$^{\rm bl}$ &14.99 $\pm$ 0.06 &15.31$\pm$ 0.16& $\cdots$ & $\cdots$ \\
V368 Aql    &13.99 $\pm$ 0.04&14.20 $\pm$ 0.08&12.15 $\pm$ 0.30& $\cdots$ \\
DO Aql$^{\rm ex}$      &15.42 $\pm$ 0.13&15.84 $\pm$ 0.27&13.07 $\pm$ 0.21& $\cdots$ \\
V476 Cyg    &15.21 $\pm$ 0.04&15.30 $\pm$ 0.08& $\cdots$ & $\cdots$ \\
RR Tel      &3.36  $\pm$ 0.11& 1.85 $\pm$ 0.01& 0.34 $\pm$ 0.01& $-$0.64 $\pm$ 0.01\\
V2467 Cyg$^{\rm ex}$   &12.53 $\pm$ 0.04&11.44 $\pm$ 0.04& 7.79 $\pm$ 0.05& 3.80 $\pm$ 0.04\\
V1974 Cyg$^{\rm ex}$   &14.99 $\pm$ 0.11&14.87 $\pm$ 0.16&12.24 $\pm$ 0.15& 8.14 $\pm$ 0.22\\
HR Del      &11.85 $\pm$ 0.02&11.69 $\pm$ 0.02&10.12 $\pm$ 0.05& 5.99 $\pm$ 0.04\\
V1330 Cyg$^{\rm ex}$   &15.11 $\pm$ 0.04&15.97 $\pm$ 0.27& $\cdots$ & $\cdots$ \\
V450 Cyg    &14.22 $\pm$ 0.03&14.38 $\pm$ 0.06& $\cdots$ & $\cdots$ \\
V2275 Cyg$^{\rm ex}$   &14.82 $\pm$ 0.07&14.90 $\pm$ 0.16& $\cdots$ & $\cdots$ \\
Q Cyg       &12.84 $\pm$ 0.02&12.81 $\pm$ 0.03& $\cdots$ & $\cdots$ \\
IV Cep$^{\rm bl,ex}$ &12.56 $\pm$ 0.07&12.49 $\pm$ 0.11& $\cdots$ & $\cdots$ \\
DI Lac      &13.33 $\pm$ 0.03&13.33 $\pm$ 0.03&12.73 $\pm$ 0.34 $\cdots$ \\
DK Lac      &14.93 $\pm$ 0.04&15.03 $\pm$ 0.12& $\cdots$ & $\cdots$ \\
V705 Cas    &14.03 $\pm$ 0.03&13.91 $\pm$ 0.04&10.88 $\pm$ 0.20&7.25 $\pm$ 0.12\\
BC Cas      &14.51 $\pm$ 0.04&14.30 $\pm$ 0.05& $\cdots$ & $\cdots$ \\
\hline
\enddata
\begin{flushleft}
{\small
$^{\rm bl}$This object suffers from blending issues.\\
$^{\rm ex}$Indicates an object where aperture photometry using IRAF
was performed to extract the listed fluxes.\\
}
\end{flushleft}
\end{deluxetable}

\begin{figure}
\centering
\centerline{{\includegraphics[width=4in]{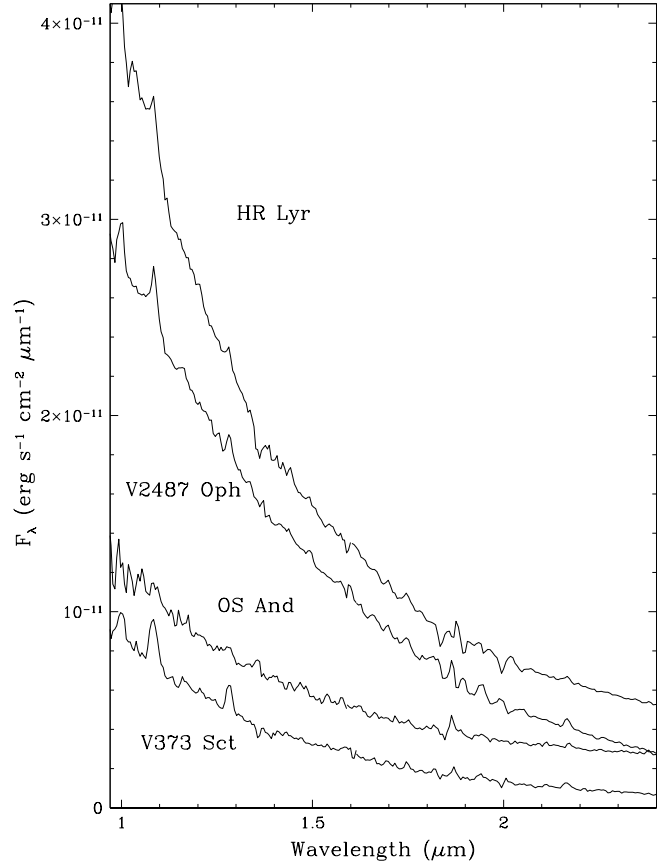}}}
\caption{The near-infrared spectra of four old classical novae obtained
with NIRC (for clarity, the spectra of HR Lyr and OS And have been offset 
vertically by 3.0 $\times$ 10$^{\rm -12}$ and 2 $\times$ 10$^{\rm -12}$ 
erg s$^{\rm -1}$ cm$^{\rm -2}$, respectively). The extrodinarily blue 
continuum of HR Lyr may indicate the presence of cyclotron emission.}
\end{figure}

\begin{figure}
\centering
\centerline{{\includegraphics[width=4in]{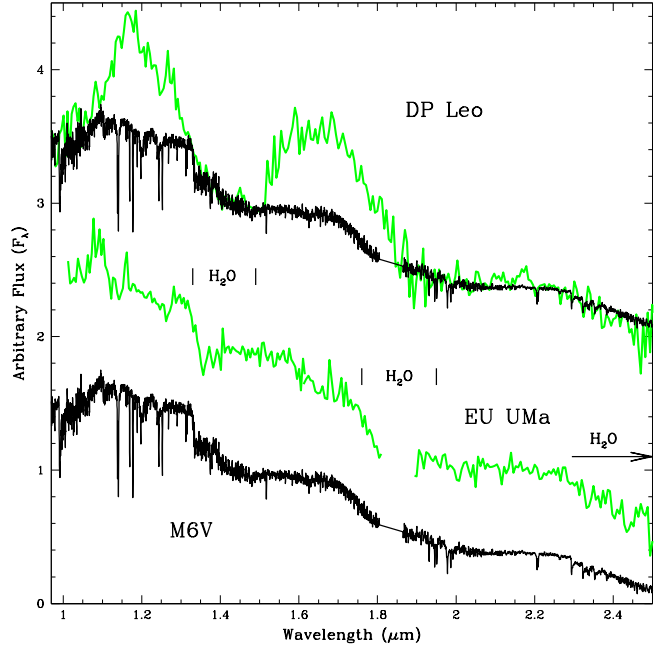}}}
\caption{The spectra of DP Leo (top) and EU UMa (middle) compared to the 
spectrum of an M6V from the IRTF Spectral Library. The match of the water 
vapor features (delineated), and the $K$-band continuum, of the M6V to DP Leo 
and EU UMa is quite good. The excess above the late-type star continuum in
DP Leo is cyclotron emission from the $n$ = 2 and $n$ = 3 harmonics of the 31 
MG field on the white dwarf primary. The spectrum of EU UMa shows little 
evidence for significant cyclotron emission. }
\end{figure}

\begin{figure}
\centering
\centerline{{\includegraphics[width=4in]{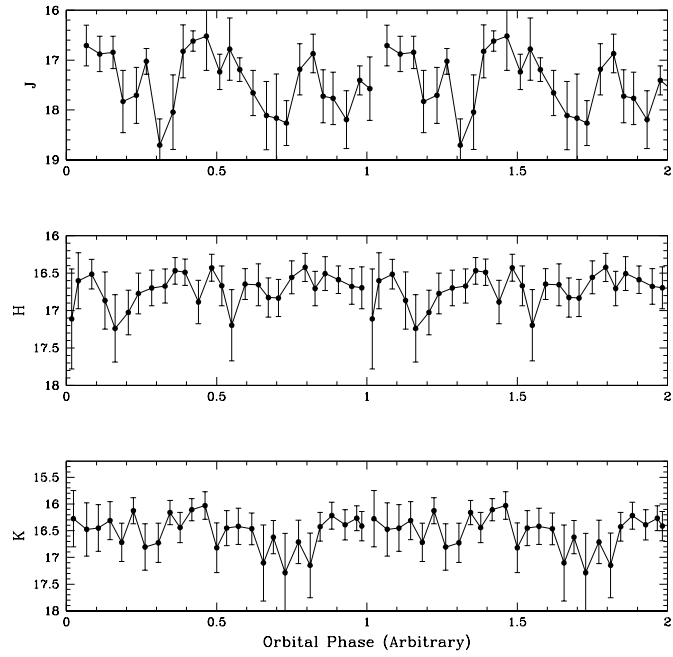}}}
\caption{The near-infrared light curve of EU UMa, showing large amplitude 
variations in the $J$-band. Such large variations are frequently seen in 
the light curves of polars, and suggests that there is cyclotron emission 
present in this system.}
\end{figure}

\begin{figure}
\centering
\includegraphics[width=4in]{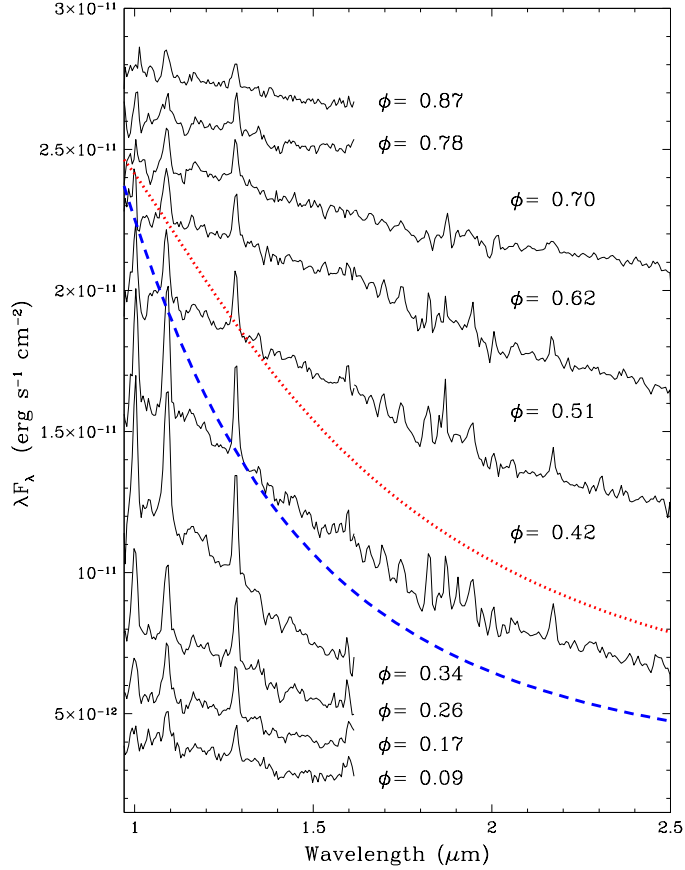}
\caption{Phase resolved spectra of V1500 Cyg. The ``JH'' spectra were 
obtained on 2006 July 7, and the ``HK'' spectra were obtained one night 
later. The HK spectra have been given small flux offsets to join smoothly 
onto the JH data set. Note that the flux has been plotted in 
$\lambda$F$_{\lambda}$ to help accentuate any cyclotron features if they 
happened to be present (see Campbell et al. 2008c for example cyclotron 
spectra plotted in this fashion). In addition, for clarity, from bottom
to top the $JHK$
spectra have been given the following offsets: 0, 0, 0, $-$0.2, 0.3, 1.1, 1.6,
1.9, 2.2, 2.4 ($\times$ 10$^{\rm -11}$ erg s$^{\rm -1}$
cm$^{\rm -2}$). The red (dotted) curve is the spectrum of a 
3000 K blackbody, while the blue (dashed) curve represents an 8800 K blackbody. 
The numbers to the right of each spectrum are the phase according to the 
ephemeris of Semeniuk et al. (1995a). At this observational epoch, the
photometric and magnetic phases (see the text describing Fig. 5b) were 
nearly identical.}
\end{figure}

\renewcommand{\thefigure}{5a}
\begin{figure}
\centering
\includegraphics[width=4in]{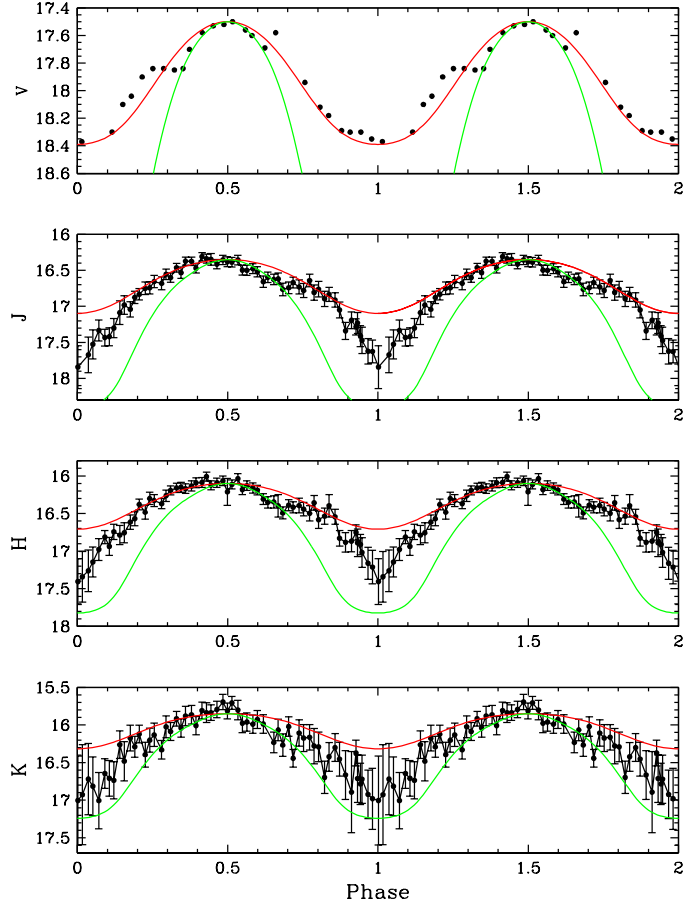}
\caption{Models for the phase-resolved $VJHK$ light curves of V1500 Cyg, 
using the parameters (T$_{\rm 1}$ = 90,000 K, T$_{\rm 2}$ = 3,000 K, and 
$i$ = 60$^{\circ}$) from Schmidt et al. (1995) in green. The red model has 
T$_{\rm 1}$ = 58,500 K, T$_{\rm 2}$ = 3,000 K, and $i$ = 30$^{\circ}$.
We have arbitrarily set the time of the deep minima seen in the photometry
to phase 0 for the light curve modeling. }
\end{figure}

\renewcommand{\thefigure}{5b}
\begin{figure}
\centering
\includegraphics[width=4in]{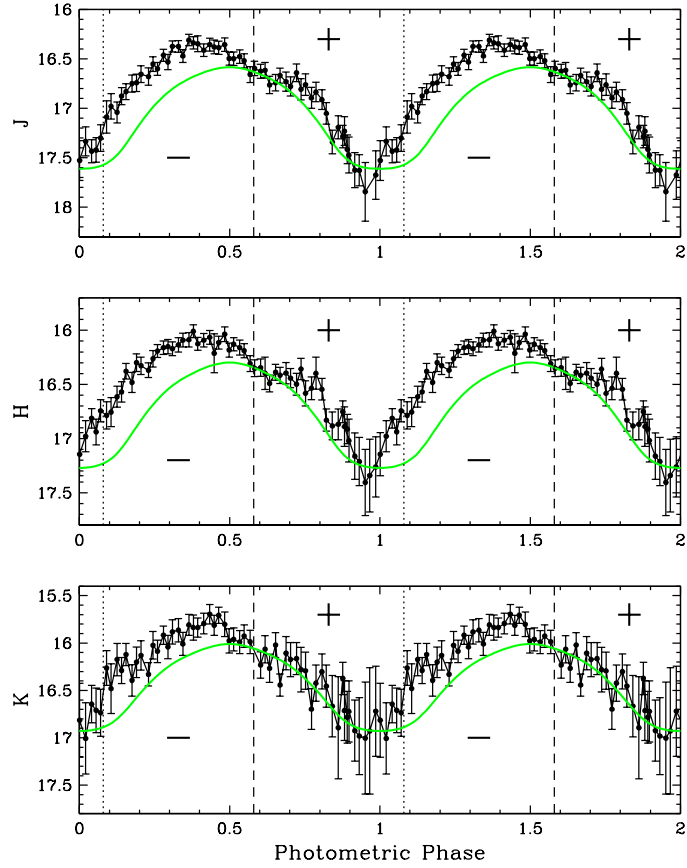}
\caption{The $JHK$ light curves of V1500 Cyg phased to the 
photometric ephemeris of V1500 Cyg by Semeniuk et al. (1995a). In the near-IR, 
the observed maxima and minima do not correspond exactly to the correct 
phases for the predicted maxima and minima. We indicate the location of the 
magnetic phase zero with a vertical dashed line, and magnetic phase 0.5 with a 
dotted vertical line. We overplot the Schmidt et al. (1995) light curve model
but have forced it to have its maximum near magnetic phase 0, and a minimum 
near photometric phase 0. The ``$+$'' and ``$-$'' signs are plotted at the 
times of maximum positive and negative circular polarizations. }
\end{figure}

\renewcommand{\thefigure}{6}
\begin{figure}
\centering
\includegraphics[width=4in]{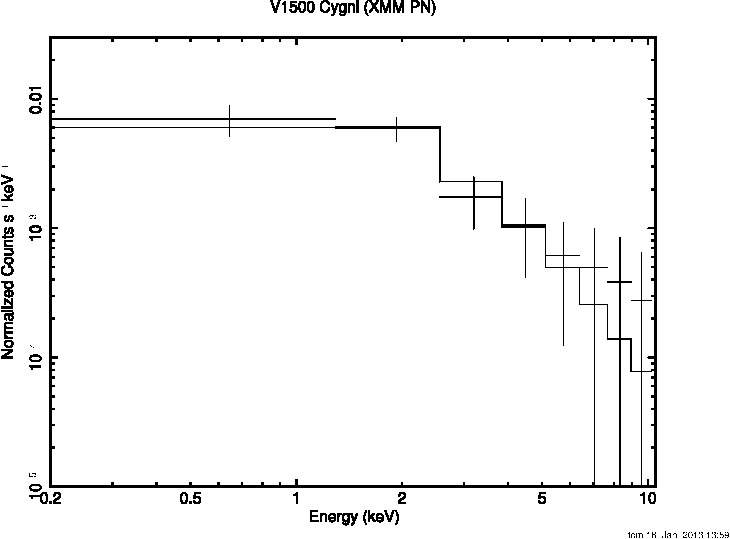}
\caption{The XMM spectrum of V1500 Cyg. It has been modeled (solid line)
using a thermal bremsstrahlung source with a temperature of $k$T = 4.2 keV,
the best-fitting single component model.}
\end{figure}

\renewcommand{\thefigure}{7}
\begin{figure}
\centering
\includegraphics[width=4in]{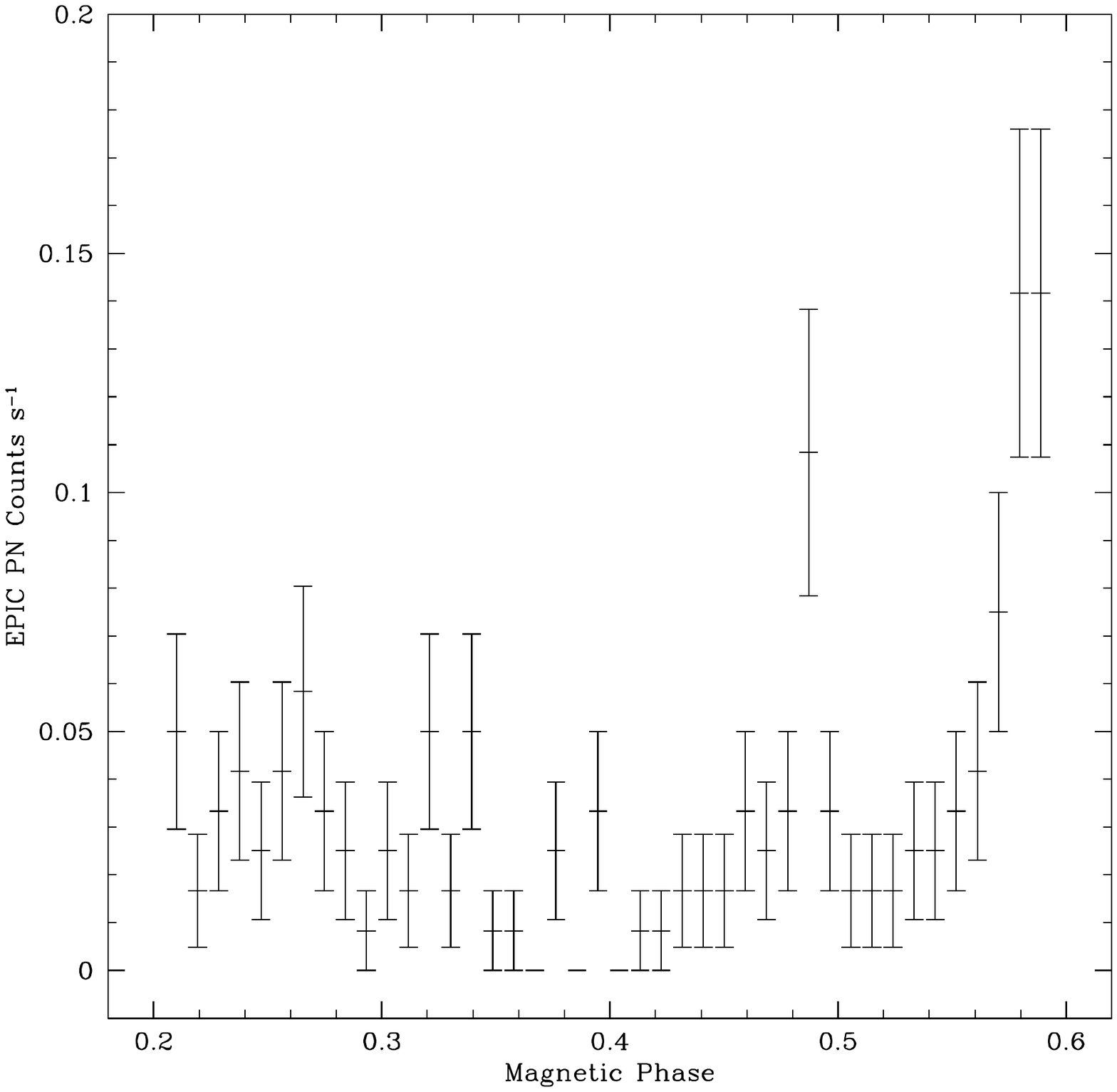}
\caption{The XMM light curve of V1500 Cyg phased to the magnetic (rotational 
``spin'' of the white dwarf) ephemeris. The rise at the end of this
light curve is expected, since the X-ray emission is directly correlated
with the optical polarization, and the strongest peak of the latter occurs
at magnetic phase 0.75. }
\end{figure}

\renewcommand{\thefigure}{8}
\begin{figure}
\centering
\includegraphics[width=4in]{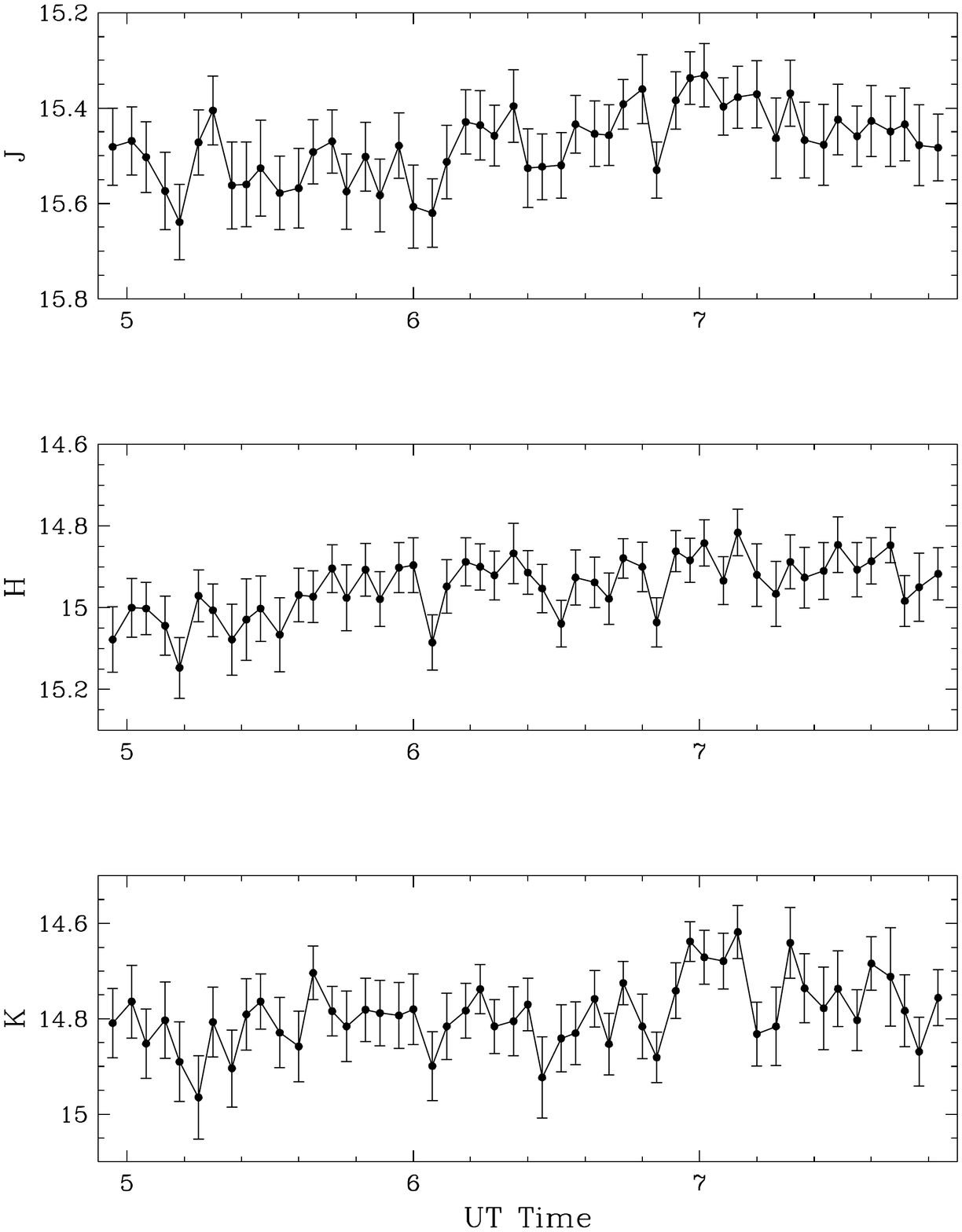}
\caption{The $JHK$ light curves of V2487 Oph obtained using SQIID. The orbital
period for this object remains unknown.}
\end{figure}

\renewcommand{\thefigure}{9}
\begin{figure}
\centering
\includegraphics[width=4in]{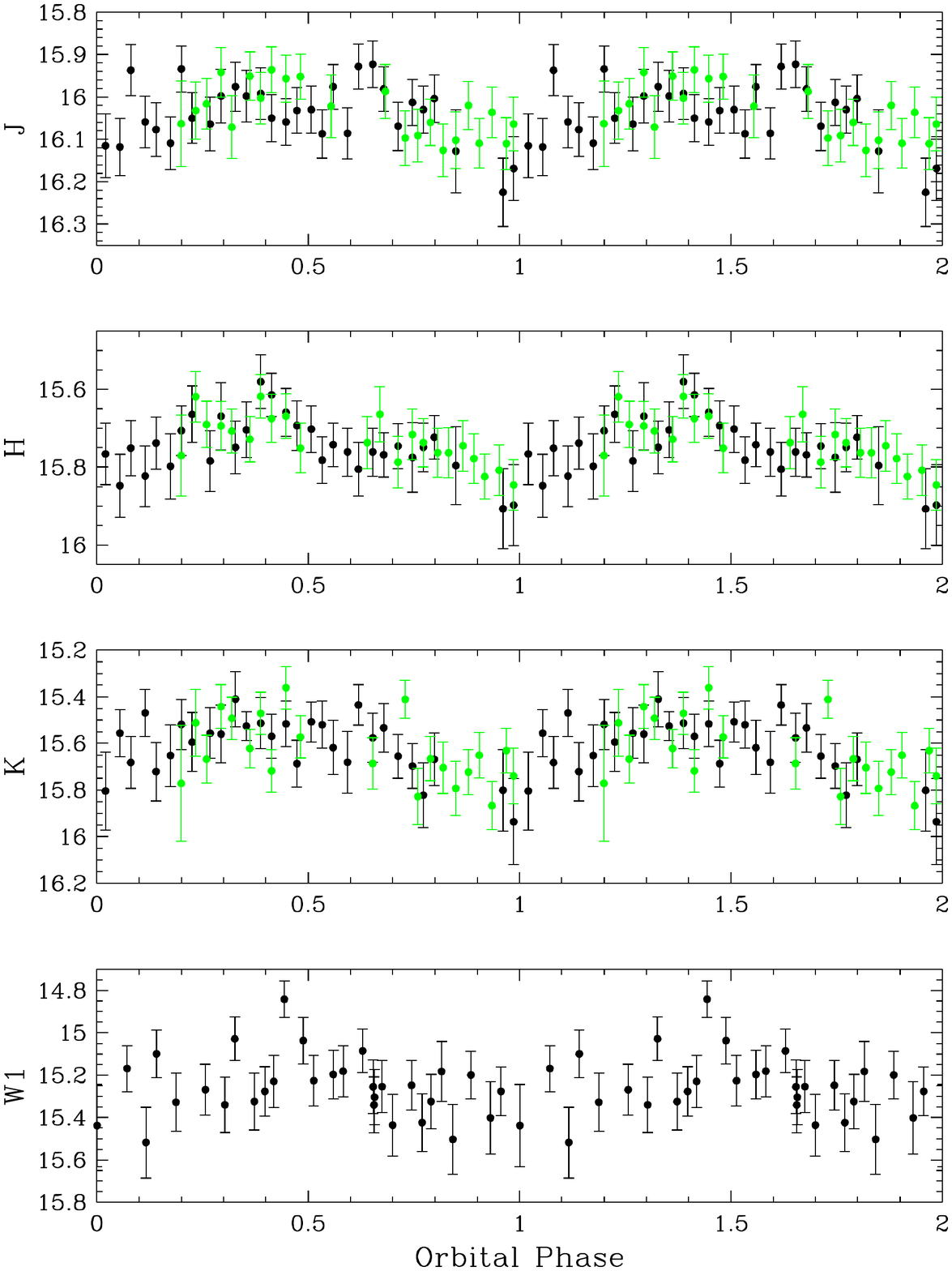}
\caption{The $JHK$ and $WISE$ W1 light curves of V1974 Cyg. These light
curves superficially resemble those of V1500 Cyg, and suggest the possibility of
significant irradiation of the secondary star, or possible cyclotron emission. 
As noted in Table 1, two separate SQIID data sets were obtained for this 
object on 2006 July 13. The black data points represent the data set with the 
05:52 UT start time, and the green data points represent the photometry with 
the start time of 09:49 UT. The phasing is from Semeniuk et al. (1995b).}
\end{figure}

\renewcommand{\thefigure}{10}
\begin{figure}
\centering
\includegraphics[width=4in]{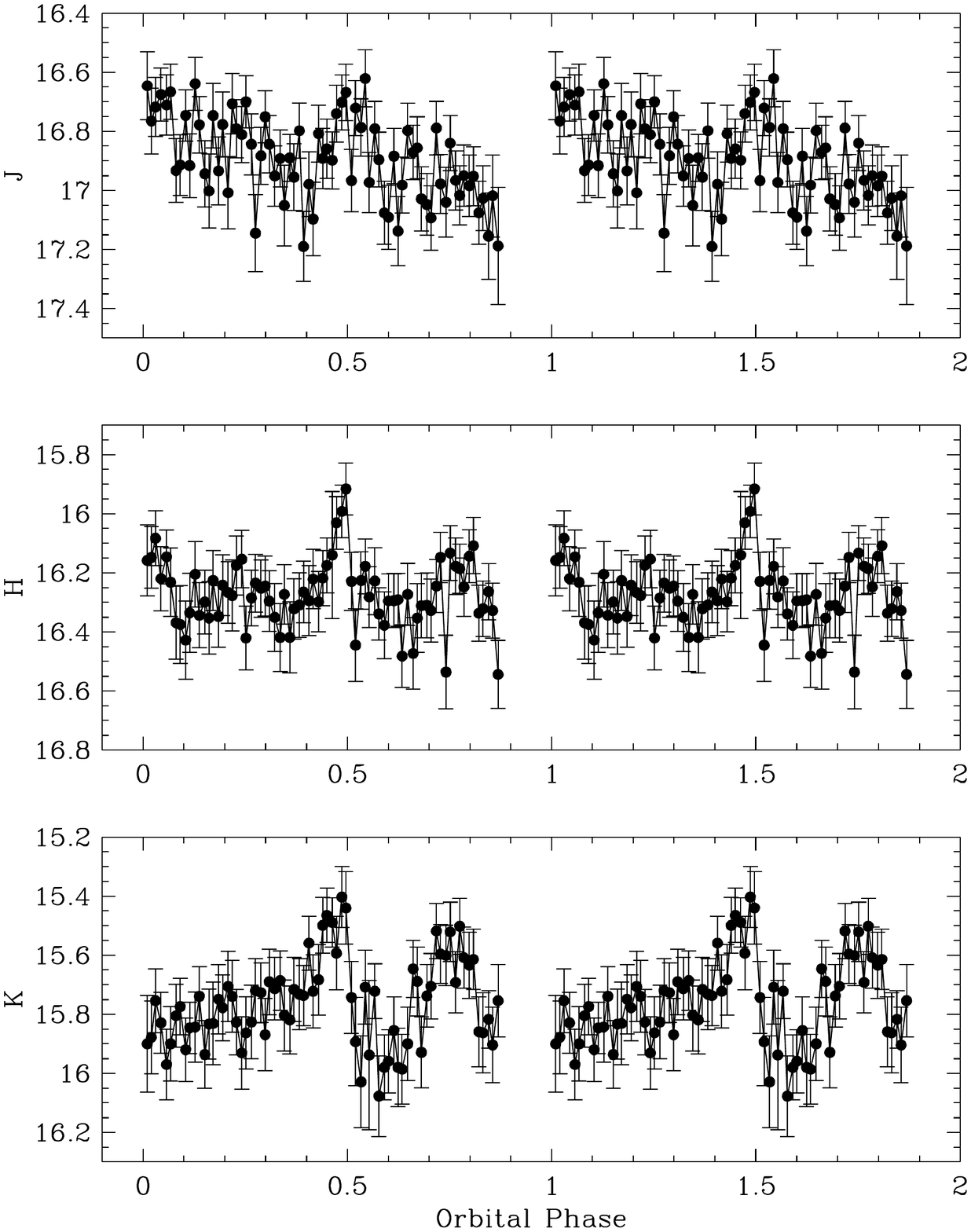}
\caption{The near-infrared light curve of V446 Her obtained using SQIID. Thorstensen
\& Taylor (2000) present a radial velocity study of this system, and develop
an ephemeris based on H$\alpha$. It is unclear how this translates into orbital
phase, so we have chosen to simply arbitrarily phase our data using their orbital
period for this object.}
\end{figure}

\renewcommand{\thefigure}{11}
\begin{figure}
\centering
\includegraphics[width=4in]{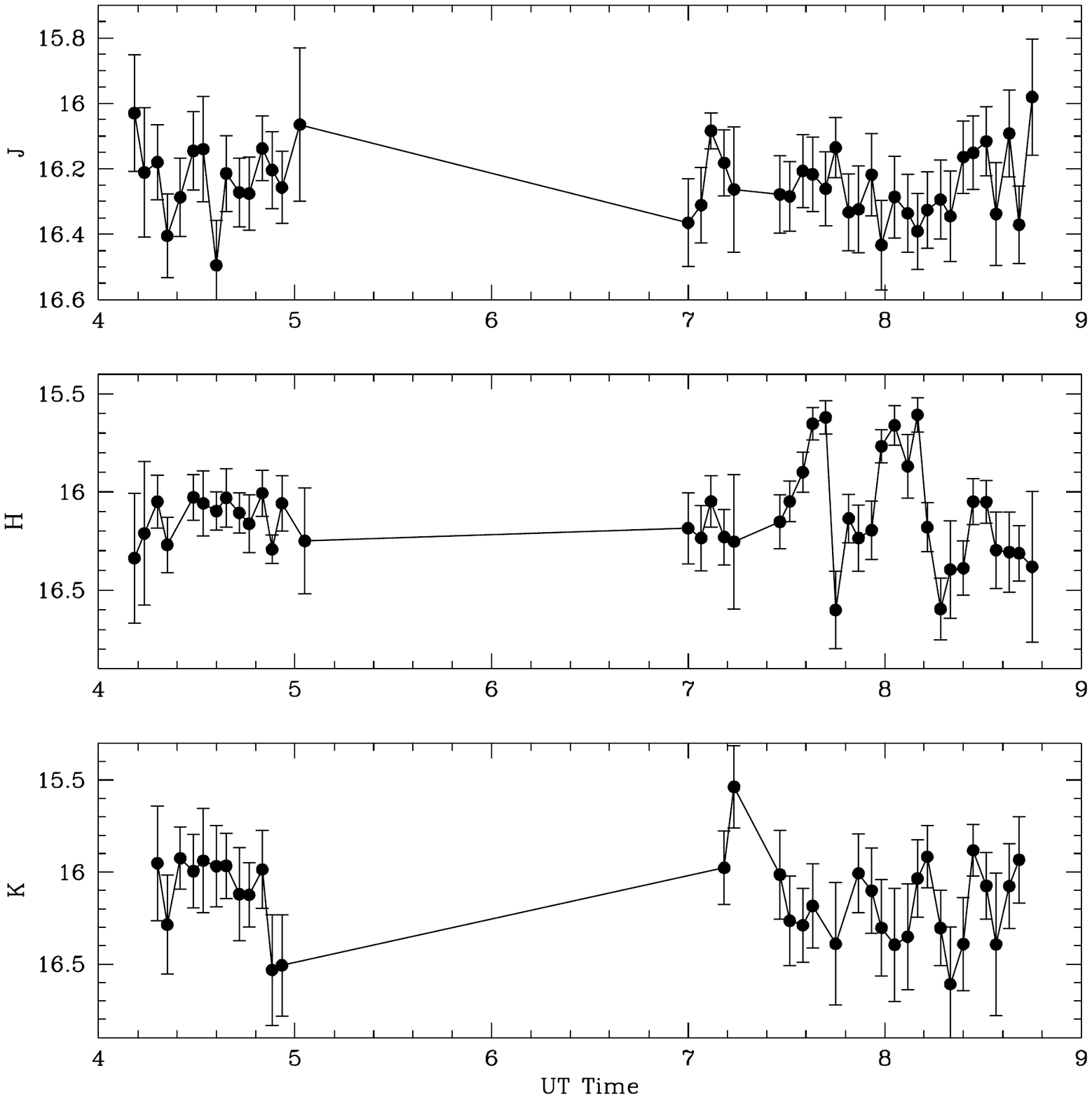}
\caption{The near-infrared light curve of QV Vul obtained using SQIID. The observations
were interrupted by thick cloud. The orbital period of QV Vul remains unknown.}
\end{figure}

\renewcommand{\thefigure}{12}
\begin{figure}
\centering
\includegraphics[width=4in]{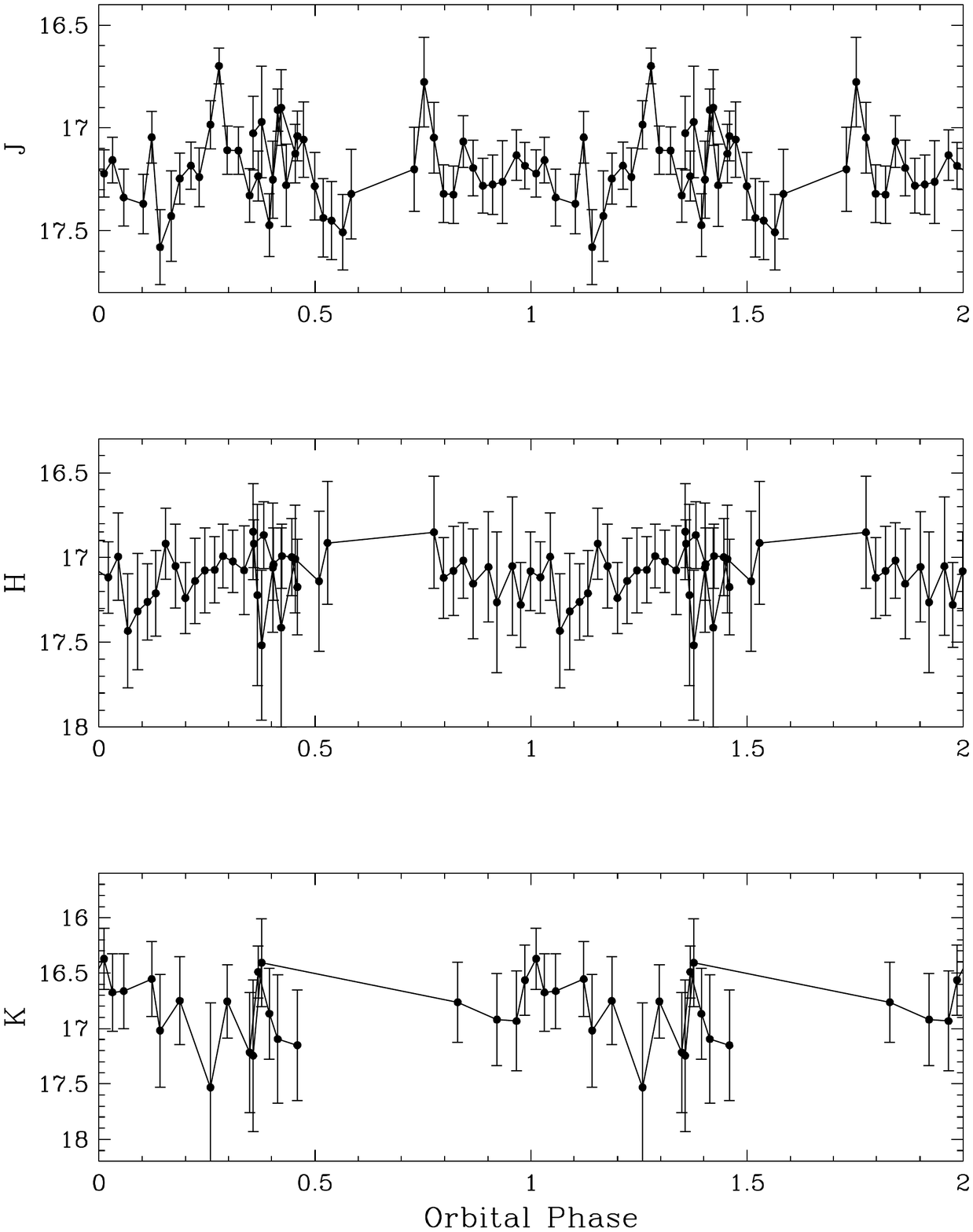}
\caption{The near-infrared light curve of V Per obtained using SQIID. We have phased
this light curve using the ephemeris by Shafter \& Abbot (1989) who showed that V Per
is an eclipsing system.}
\end{figure}

\renewcommand{\thefigure}{13}
\begin{figure}
\centering
\includegraphics[width=4in]{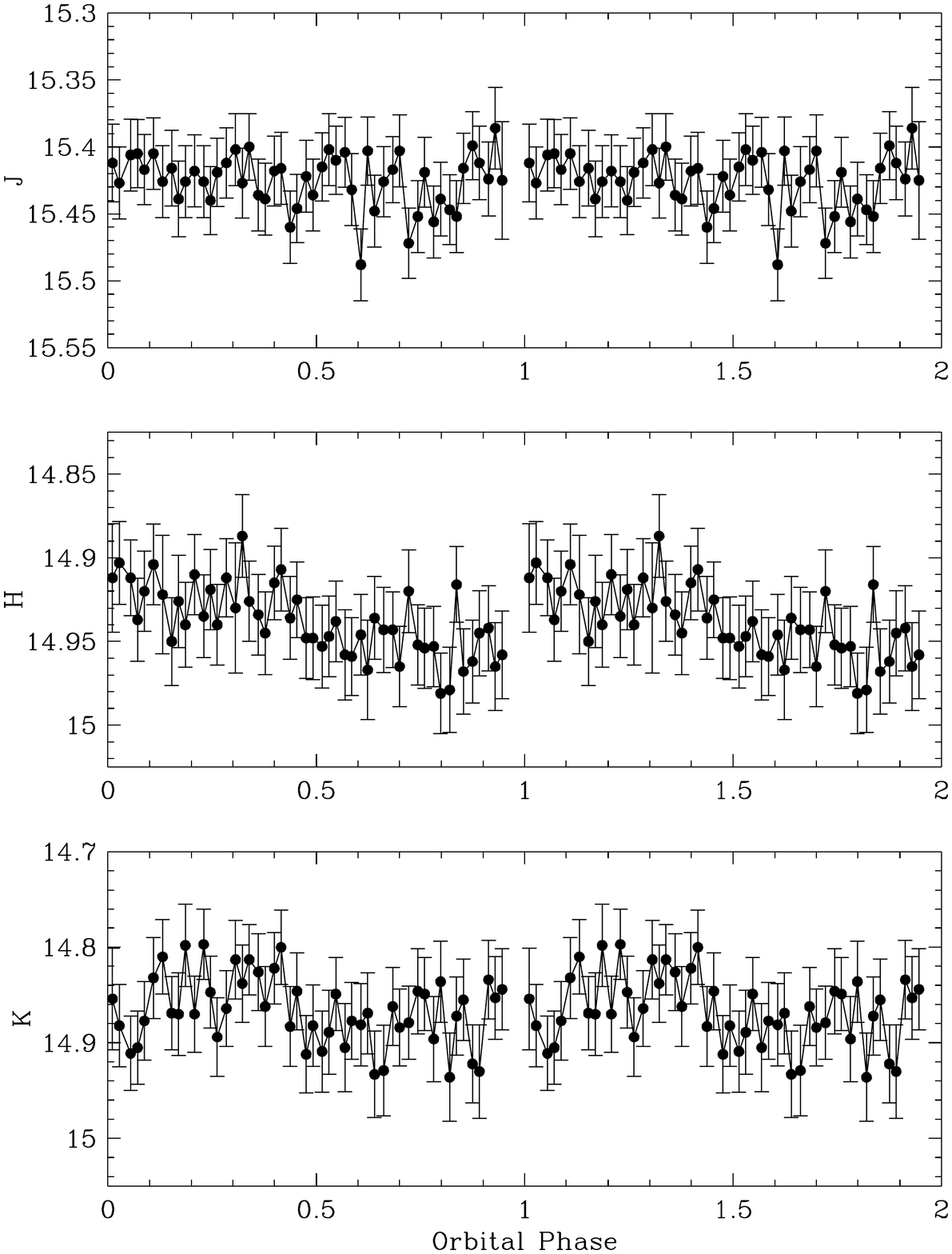}
\caption{The near-infrared light curve CP Lac obtained using SQIID. While the
orbital period of CP Lac is known, an orbital ephemeris has yet to be established. }
\end{figure}

\renewcommand{\thefigure}{14}
\begin{figure}
\centering
\includegraphics[width=4in]{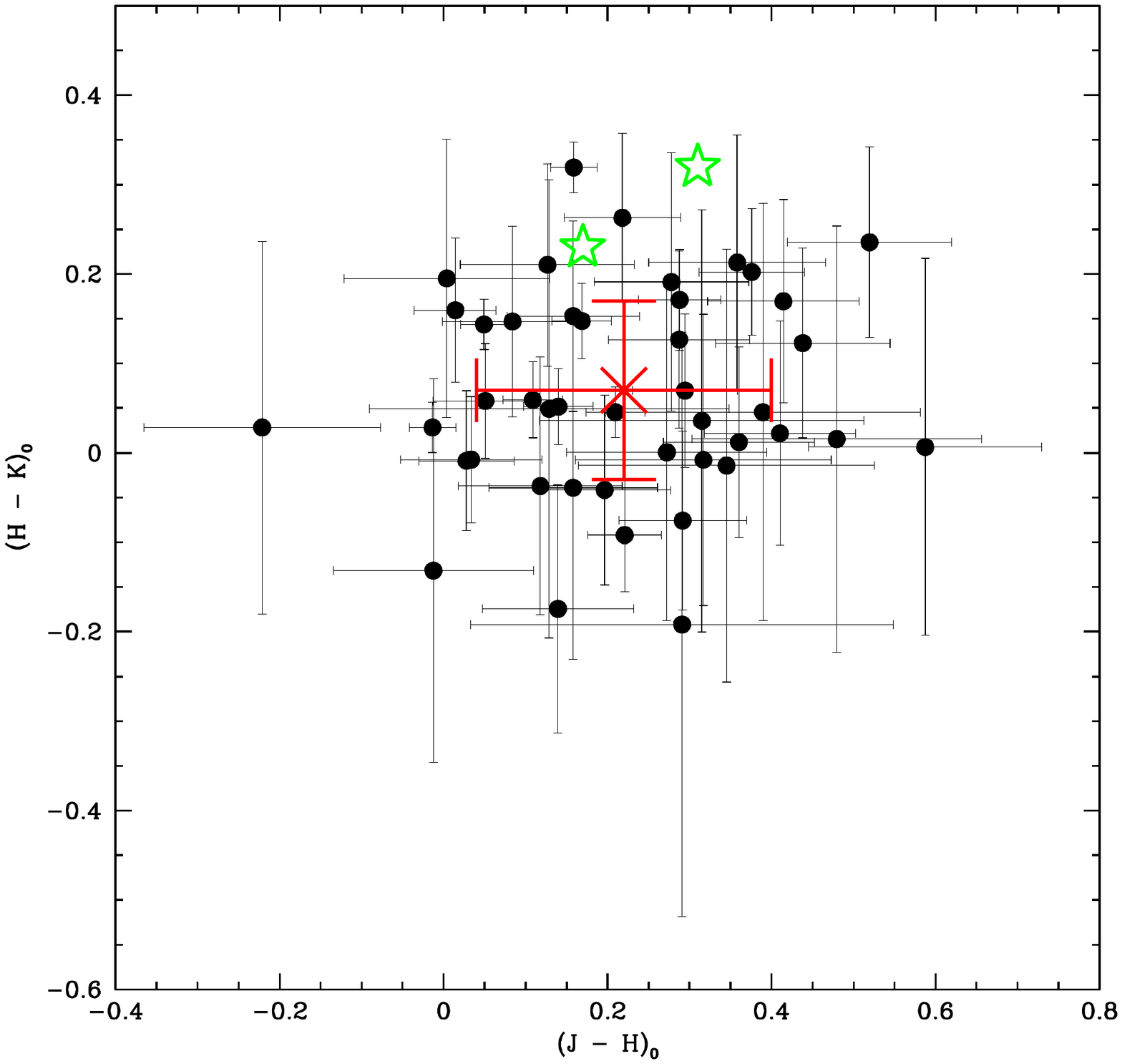}
\caption{The de-reddened 2MASS color-color plot for the CNe in Table 3 with pre-existing
estimates of their visual extinctions. The mean colors for the sample is denoted with
an ``$\times$''. The colors of V1500 Cyg at minimum and maximum are indicated by
the star symbols.}
\end{figure}

\renewcommand{\thefigure}{15}
\begin{figure}
\centering
\includegraphics[width=4in]{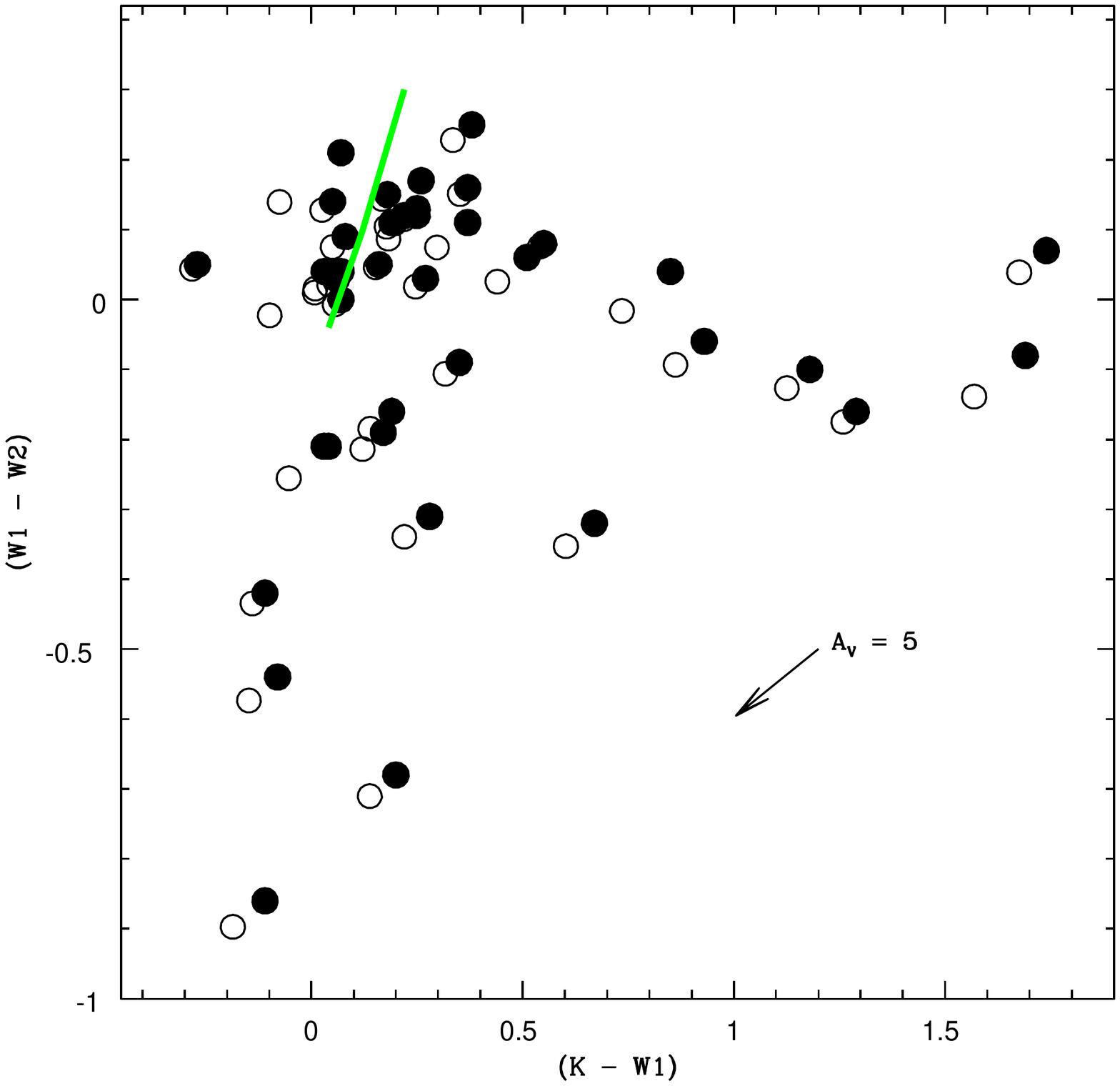}
\caption{The (W1 $-$ W2) vs. ($K_{\rm 2MASS}$ $-$ W1)
color-color plot for the CNe in Table 4. The solid circles are the observed
data, the open circles have been de-reddened using the extinction values
listed in Table 3, and with relationships from Yuan et al. (2013). The
extinction vector is plotted, as is the locus (green) of main sequence stars in
this diagram (increasing from F8V to M6V).  }
\end{figure}

\renewcommand{\thefigure}{16}
\begin{figure}
\centering
\includegraphics[width=4in]{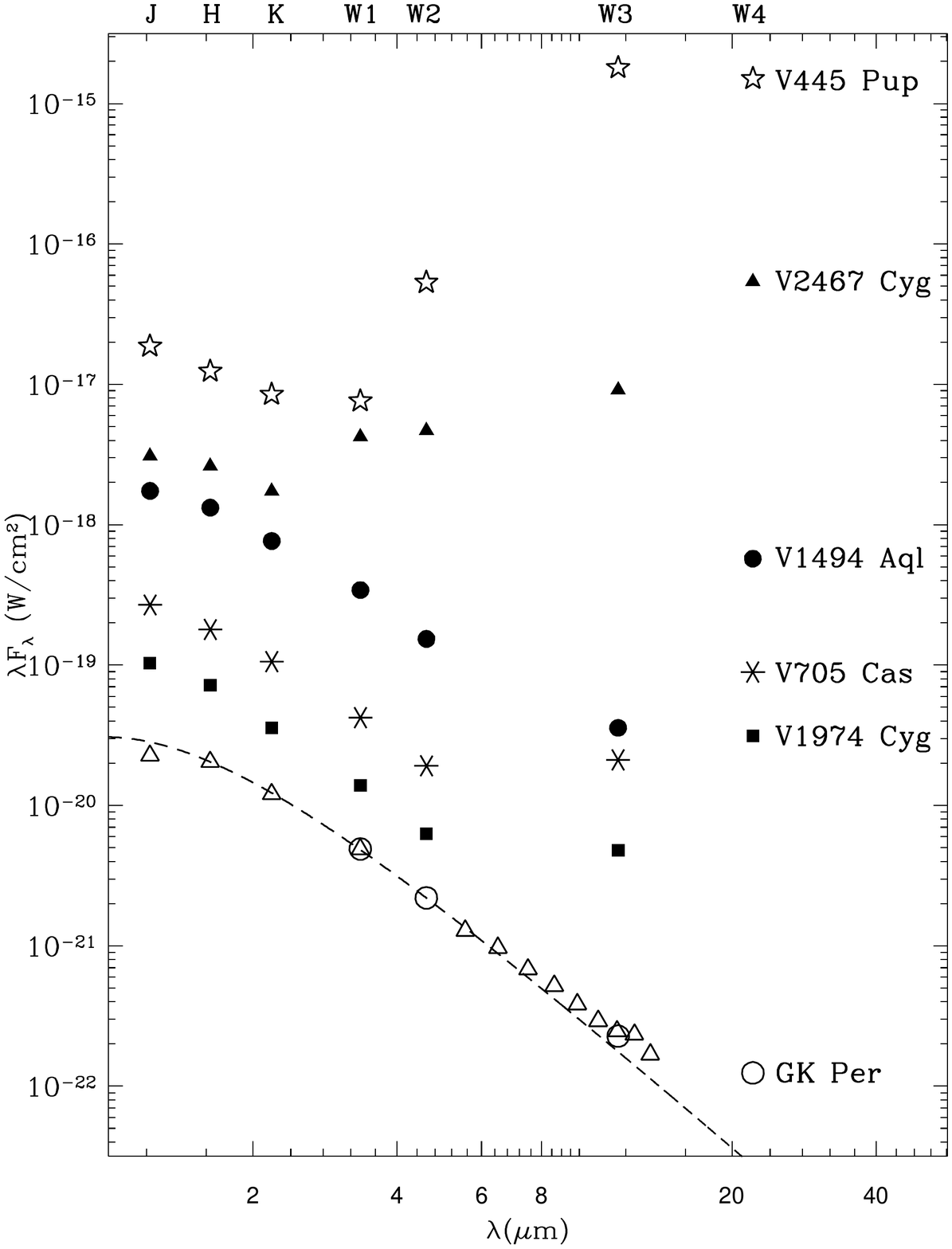}
\caption{The SEDs for six CNe that were detected in the $WISE$ W4 band.
These SEDs have been scaled for clarity. The dashed line fit to the optical
and infrared photometry of GK Per (open triangles from Harrison et al. 2007,
open circles are the $WISE$ data) is a blackbody with a temperature of 
4200 K. In all cases, the error bars on the photometry are smaller than
the sizes of the symbols.}
\end{figure}
\clearpage
\renewcommand{\thefigure}{17}
\begin{figure}
\centering
\includegraphics[width=4in.]{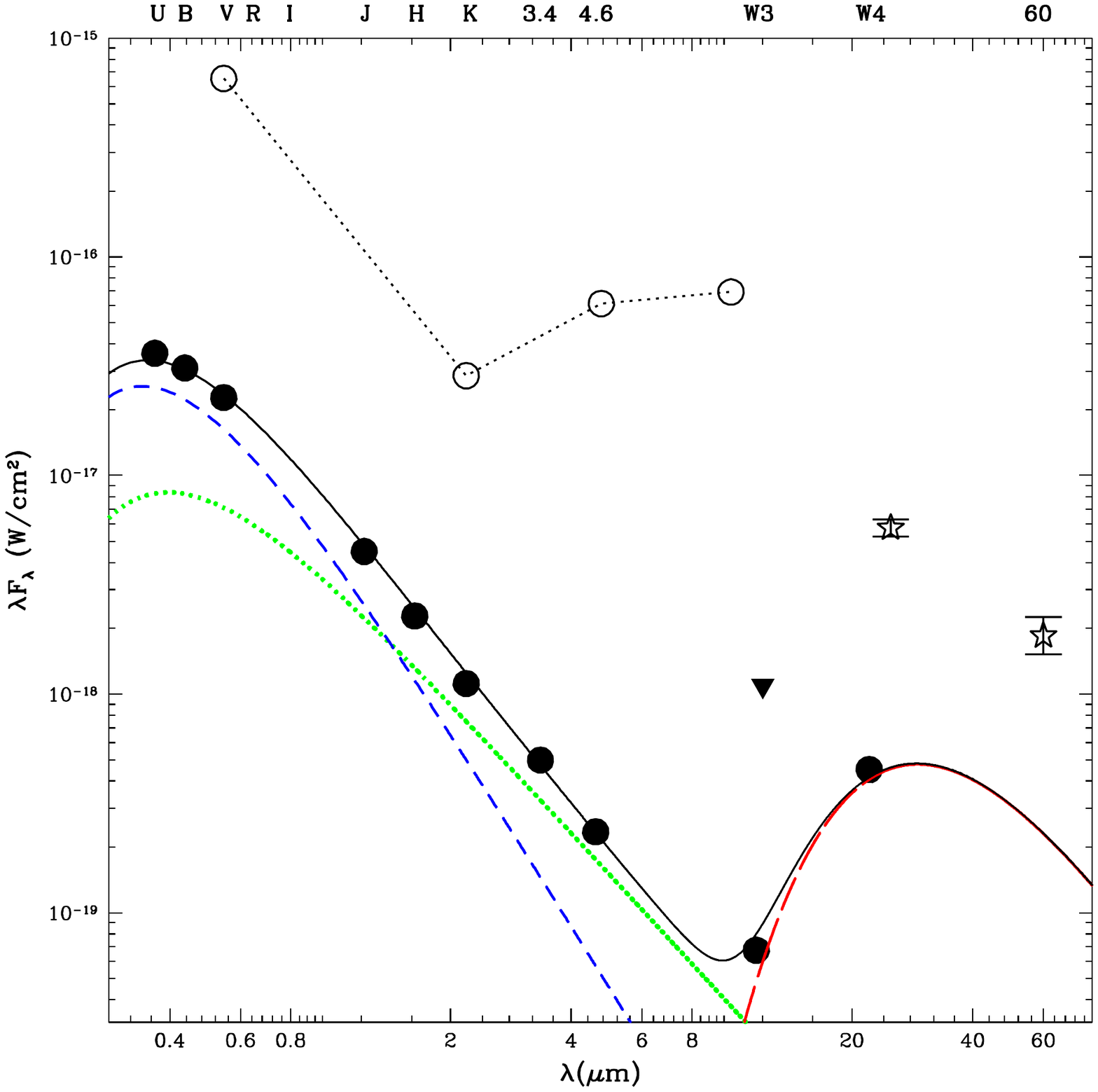}
\caption{The SED for HR Del. We have modeled this SED (solid black line)
as the sum of a hot blackbody source (65000 K, blue short-dashed line,
see Moraes \& Diaz 2009), a free-free component (green dotted line), and 
a cool blackbody (126 K, red long-dashed line), all reddened by 
A$_{\rm V}$ = 0.49 mag. The open circles are the second epoch observations
by GKL. The star symbols with error bars (and 12 $\mu$m upper limit) are the 
$IRAS$ observations. The errors on the photometry are smaller than
the symbol size for those data without error bars.}
\end{figure}

\end{document}